\documentclass[12pt]{article}
\pdfoutput=1

\usepackage{jheppub}
\usepackage{amssymb,amsfonts,amsmath,amsthm, enumerate}
\usepackage{mathrsfs}
\usepackage{bbold}
\usepackage{tikz}
\usetikzlibrary{shapes.geometric, arrows}
\usepackage{comment}
\usepackage{tikz}
\usepackage{hyperref}

\usepackage{arydshln}

\usepackage{calc}

\usepackage{accents}

\makeatletter
\newcommand{\Vast}{\bBigg@{4.75}}
\makeatother

\newcommand{\be}{\begin{equation}}
\newcommand{\ee}{\end{equation}}
\newcommand{\bea}{\begin{eqnarray}}
\newcommand{\eea}{\end{eqnarray}}

\newcommand{\CA}{\mathcal{A}}
\newcommand{\CB}{\mathcal{B}}
\newcommand{\CC}{\mathcal{C}}
\newcommand{\CD}{\mathcal{D}}
\newcommand{\CF}{\mathcal{F}}
\newcommand{\CG}{\mathcal{G}}
\newcommand{\CH}{\mathcal{H}}

\newcommand{\CJ}{\mathcal{J}}

\newcommand{\CM}{\mathcal{M}}

\newcommand{\CT}{\mathcal{T}}

\newcommand{\CV}{\mathcal{V}}

\newcommand{\lr}{\left (}
\newcommand{\rr}{\right )}
\newcommand{\ls}{\left [}
\newcommand{\rs}{\right ]}
\newcommand{\lc}{\left \{}
\newcommand{\rc}{\right \}}

\newcommand\qt\tau

\newcommand{\co}{\varphi}

\newcommand{\p}{\partial}

\renewcommand{\tilde}[1]{\widetilde{#1}}

\makeatletter
\renewcommand{\@seccntformat}[1]{\csname the#1\endcsname.\,\,}
\makeatother
\allowdisplaybreaks

\let \savenumberline \numberline
\def \numberline#1{\savenumberline{#1.}}

\makeatletter
\def\@fpheader{\relax}
\makeatother

\def\bea{\begin{eqnarray}}
\def\eea{\end{eqnarray}}

\usetikzlibrary{shapes,arrows,chains}
\usetikzlibrary{decorations.markings}
\usetikzlibrary{decorations.pathmorphing}
\usetikzlibrary{shapes.multipart}
\tikzset{snake it/.style={decorate, decoration=snake}}

\title{\ \vspace{1.6cm} \\
\scalebox{0.85}{Torsional Deformation of Nonrelativistic String Theory}}
\author{Ziqi Yan}
\emailAdd{ziqi.yan@su.se}
\affiliation{
Nordita, KTH Royal Institute of Technology and Stockholm University\\
Hannes Alfv\'{e}ns v\"{a}g 12, SE-106 91 Stockholm, Sweden}
\abstract{Nonrelativistic string theory is a self-contained corner of string theory, with its string spectrum enjoying a Galilean-invariant dispersion relation. This theory is unitary and ultraviolet complete, and can be studied from first principles. In these notes, we focus on the bosonic closed string sector. In curved spacetime, nonrelativistic string theory is defined by a renormalizable quantum nonlinear sigma model in background fields, following certain symmetry principles that disallow any deformation towards relativistic string theory. We review previous proposals of such symmetry principles and propose a modified version that might be useful for supersymmetrizations. The appropriate target-space geometry determined by these local spacetime symmetries is string Newton-Cartan geometry. This geometry is equipped with a two-dimensional foliation structure that is restricted by torsional constraints. Breaking the symmetries that give rise to such torsional constraints in the target space will in general generate quantum corrections to a marginal deformation in the worldsheet quantum field theory. Such a deformation induces a renormalization group flow towards sigma models that describe relativistic strings.}%

\begin{document}

\maketitle
\vfill\eject

\section{Introduction and Conclusions}

In recent years, there has been a growing interest in non-Lorentzian geometries that arise in string theory. One such corner of string theory that enjoys nonrelativistic symmetries was put forward more than twenty years ago, under the name \emph{nonrelativistic string theory} \cite{Klebanov:2000pp, Gomis:2000bd, Danielsson:2000gi}. This theory is a unitary and ultraviolet (UV) complete quantum gravity theory with a Galilean-invariant string spectrum. Via T-duality \cite{Gomis:2000bd, Danielsson:2000gi, Bergshoeff:2018yvt, Gomis:2020izd}, it gives a microscopic definition of string theory in the discrete light cone quantization (DLCQ), which is relevant for Matrix string theory \cite{Banks:1996my, Dijkgraaf:1997vv, Motl:1997th}. When we consider open strings in nonrelativistic string theory, Galilean-invariant Yang-Mills theories also emerge \cite{Gomis:2020fui}. 

In flat spacetime, nonrelativistic string theory is defined by a two-dimensional relativistic quantum field theory (QFT) with a nonrelativistic global symmetry group. In addition to regular worldsheet fields that parametrize the target space coordinates, there are also a pair of one-form worldsheet fields in the formalism of nonrelativistic string theory, namely, a (1,0)-form $\lambda$ and a (0,1)-form $\bar{\lambda}$\,. These one-form fields are related to dual coordinates that are conjugate to string windings, and they are responsible for the consistency and salient features in nonrelativistic string theory. For example, $\lambda$ and $\bar{\lambda}$ play the role of a Lagrange multiplier in the string sigma model and select a two-dimensional longitudinal sector in spacetime. This longitudinal sector is defined by a pair of worldsheet fields that form the spacetime lightcone coordinates. The remaining transverse spacetime directions are in form the same as in relativistic string theory. The longitudinal and transverse sectors are related to each other by stringy Galilei boosts, under which $\lambda$ and $\bar{\lambda}$ transform nontrivially.  

Interactions in the two-dimensional worldsheet QFT are induced by inserting vertex operators, from which the spacetime geometry emerges as background fields. If the only requirement is worldsheet conformal symmetry, then a marginal and classically conformal operator $\lambda \bar{\lambda}$ will be included in the spectrum. This $\lambda \bar{\lambda}$ operator deforms the sigma model towards the relativistic string sigma models \cite{Gomis:2000bd, Danielsson:2000mu, Gomis:2019zyu}. In the literature, depending on the purposes of the studies, there are two different perspectives on how this deformation term should be treated, which we detail below:

\vspace{3mm}

\noindent $\bullet$ \emph{Nonrelativistic strings from a limit.} 
The first perspective seeks a nonsingular limit of spacetime (super)gravity that exhibits nonrelativistic behaviors \cite{Harmark:2017rpg, Harmark:2018cdl, Gallegos:2019icg, Harmark:2019upf, Bergshoeff:2021bmc}. From the worldsheet point of view, this requires studying the renormalization group (RG) flow of the worldsheet QFT with all possible vertex operators. In the presence of the $\lambda\bar{\lambda}$ operator, the worldsheet QFT describes relativistic strings in background fields, where the spacetime geometry is Lorentzian and parametrized in a frame of choice with a two-dimensional foliation. This foliation structure does not persist unless additional geometric constraints are imposed. The beta-functionals of the background fields are reparametrizations of the standard ones for the metric, $B$-field, and dilaton couplings in relativistic string sigma models. Requiring Weyl invariance at the quantum level sets these reparametrized beta-functionals to zero, and leads to the spacetime (super)gravity equations of motion. At the conformal fixed point, one can fine tune the coupling (that we refer to as $U$) associated with the $\lambda\bar{\lambda}$ operator to zero such that a nonrelativistic solution to the relativistic spacetime equations of motion is obtained. The resulting geometry is non-Lorentzian and non-Riemannian, equipped with a two-dimensional foliation restricted by certain geometric constraints that arise from spacetime dynamics. This geometry is referred to as (torsional) string Newton-Cartan geometry in the literature.~\footnote{\,String Newton-Cartan geometry with the zero torsion was discussed in \cite{Andringa:2012uz, Bergshoeff:2018yvt, Bergshoeff:2019pij}. Generalizations to the torsional case were later introduced in \cite{Harmark:2019upf, Gallegos:2020egk, Bergshoeff:2021bmc}.} 

Even though we tuned the physical value of $U$ to zero at the conformal fixed point, the beta-functional of $U$ still gives rise a nontrivial equation of motion. Together with other spacetime equations of motion, the vanishing beta-functional of $U$ gives rise to torsional constraints that restrict the foliation structure. Since the quantum corrections to the $\lambda\bar{\lambda}$ operator are generated by torsions that distort the foliation in string Newton-Cartan geometry, we refer to the $\lambda\bar{\lambda}$ operator as the \emph{torsional deformation}. In this picture, the torsional constraints are determined by solving the spacetime equations of motion order by order in the Regge slope $\alpha'$. Therefore, a $\lambda\bar{\lambda}$ counterterm has to be included in the worldsheet QFT in order to determine the spacetime dynamics at each loop order. In this sense, the resulting nonrelativistic corner at $U = 0$ should be treated as a limit of relativistic string theory.

In the limiting procedure, it is possible that, at low orders in $\alpha'$, the foliation constraints determined by the spacetime dynamics are not strong enough for the beta-functionals to vanish at higher-loop orders. If this happens, there will be nonrelativistic solutions that (i) solve the spacetime equations of motion determined by dynamics at low orders in $\alpha'$, (ii) but cannot be extended to higher orders unless a small $U$ is included. This nonzero $U$ will deform the theory towards the relativistic regime.~\footnote{\,See more below \eqref{eq:dt0} in \S\ref{sec:bftd}.}

\vspace{3mm}

\noindent $\bullet$ \emph{Nonrelativistic string theory from first principles.} In the other perspective, nonrelativistic string theory is defined by a renormalizable QFT without the torsional deformation \cite{Bergshoeff:2018yvt, Bergshoeff:2019pij, Gomis:2019zyu, Yan:2019xsf}. This requirement is stronger than fine tuning the physical value of $U$ to be zero: the local counterterm associated with the $\lambda\bar{\lambda}$ operator is also excluded now. This is achieved by evoking a symmetry principle that forbids the torsional deformation from being generated at the quantum level. The target space geometry is determined by the required worldsheet symmetries, which generate the spacetime gauge transformations. In this way, nonrelativistic string theory is defined by a renormalizable worldsheet QFT that satisfies certain nonrelativistic target-space symmetries acting on worldsheet fields. This is in analogy with that relativistic string theory is defined by a two-dimensional QFT invariant under the target-space (gauged) Poincar\'{e} symmetry acting on worldsheet fields. If such a symmetry principle exists, this nonrelativistic corner will define a full-fledged string theory on the same footing as relativistic string theory, at least when the spacetime effective field theories (EFTs) are concerned.~\footnote{\,Statements in these notes for background geometries are only applicable to the zero-winding sector. When windings are included, the situation is more complex and beyond these notes.} In this way, nonrelativistic string theory can be studied from first principles, independent of its embedding in any larger framework.

In \cite{Andringa:2012uz, Bergshoeff:2018yvt}, a symmetry principle that defines nonrelativistic string sigma models is proposed, where the transverse translations and string Galilei boosts commute into a noncentral extension called $Z_A$\,, with $A$ the index of the longitudinal sector in spacetime. This noncentral extension is realized in the string sigma model as a symmetry transformation that imposes a zero-torsion condition on the longitudinal Vielbein field, before any quantum calculation is performed. This eliminates any torsion in the spacetime foliation structure. It has been shown in \cite{Gomis:2019zyu, Yan:2019xsf} that the zero-torsion condition leads to a nonrenormalization theorem: the torsional deformation $\lambda\bar{\lambda}$ is not generated at all loops. Therefore, the symmetries realized in \cite{Andringa:2012uz, Bergshoeff:2018yvt} define a notion of nonrelativistic string theory by a renormalizable worldsheet QFT.

It is recently suggested in \cite{Bergshoeff:2021bmc} that the zero-torsion condition imposed by the $Z_A$ symmetry in the worldsheet QFT might be too strong for supersymmetrizations. Moreover, it is shown in \cite{Gallegos:2019icg, Yan:2019xsf, Gomis:2020fui} that a weaker version of the torsional constraints already suffice to protect the $\lambda \bar{\lambda}$ operator from being generated by quantum corrections at the lowest loop order. These suggest that there might exist a modified symmetry principle that imposes a weaker torsional constraint in nonrelativistic string sigma models. Indeed, as we will detail in \S\ref{sec:netc}, it is possible to consistently preserve only half of the $Z_A$ transformation in the symmetry algebra. We will show that imposing the symmetry generated by $Z_0 + Z_1$ on the sigma model leads to the torsional constraints that exactly match the ones found in \cite{Bergshoeff:2021bmc} by requiring consistency with supersymmetry. We will provide a Feynman diagram argument to confirm that the halved $Z_A$ symmetry prohibits the $\lambda \bar{\lambda}$ operator from being generated by quantum corrections at all loops. From the first perspective, the torsional constraints from the halved $Z_A$ symmetry solve the vanishing beta-functional associated with the $\lambda \bar{\lambda}$ operator at $U=0$\,, with all orders in $\alpha'$ included. Therefore, the two perspectives converge in the nonperturbative regime.

\vspace{3mm}

In these notes, we discuss different symmetry principles under which nonrelativistic string theories are defined from first principles, focusing on aspects of emergent spacetime geometries. We will also frequently refer to the first perspective that treats nonrelativistic strings as a limit of relativistic string theory; this will provide useful intuition for understanding the underlying symmetries. 

These notes take the following structure. We start with a review of nonrelativistic string theory in \S\ref{sec:review}, where we first focus on the free theory and then turn on vertex operators that induce string interactions. Along the way, we develop a T-dual interpretation for the one-form fields in \S\ref{sec:ofdc}; using this, we study in \S\ref{sec:OPE} a formalism in which operator product expansions can be evaluated in a compact way. In \S\ref{sec:dtrst}, we identify the vertex operator that gives rise to the torsional deformation that drives the theory towards relativistic string theory. In \S\ref{sec:nscb}, we consider sigma models in curved spacetime. In \S\ref{sec:ssm}, we study the torsional deformation in string sigma models with arbitrary background fields. In \S\ref{sec:bftd}, we derive the beta-functionals that arise from fine-tuning the torsional deformation to zero in the relativistic beta-functionals, using the results given in \cite{Bergshoeff:2021bmc}. Finally, in \S\ref{sec:netc}, we discuss different spacetime gauge symmetries that act on worldsheet fields and apply them to define a renormalizable worldsheet QFT, using which nonrelativistic string theories can be studied from first principles. In addition, we give a list of appendices on relevant topics, including T-duality transformations in Appendix \ref{app:Tdual} and the Hamiltonian formalism in Appendix \ref{app:Hfgm}. In Appendix \ref{app:nonre}, we use a Feynman-diagram argument to show that the $\lambda\bar{\lambda}$ operator is not generated at all loops in presence of the halved $Z_A$ symmetry. In Appendix \ref{app:sa}, we summarize different symmetry algebras that are used throughout the notes.

\vspace{3mm}

\emph{Note Added:} In the final stage of this work, we heard from Leo Bidussi, Troels Harmark, Jelle Hartong, Niels A. Obers, and Gerben Oling of their study of classical sigma models in a string Newton-Cartan geometry without geometric constraints \cite{nsncgnrs}.

\section{Nonrelativistic Strings in Flat Spacetime} \label{sec:review}

We start with collecting ingredients in nonrelativistic string theory that will be essential in later discussions. Nonrelativistic string theory is defined on a two-dimensional Riemann surface $\Sigma$ that acts as the worldsheet, parametrized by $\sigma^\alpha = (\tau, \sigma)$ and equipped with a worldsheet metric $h_{\alpha\beta}$\,. The imaginary time $\tau$ is related to the real time $t$ via $\tau = i t$\,. The worldsheet $\Sigma$ is mapped to a foliated spacetime manifold $\CM$ by the worldsheet coordinates $X^\mu (\sigma^\alpha) = (X^A, X^{A'})$\,, with $A = 0\,,1$ and $A' = 2, \cdots\!, d-1$\,. We refer to the two-dimensional foliation with coordinates $X^A$ as the longitudinal sector and the leaves with coordinates $X^{A'}$ as the transverse sector. Nonrelativistic string theory in flat spacetime is defined by the following free action in imaginary time \cite{Bergshoeff:2018yvt}:
\begin{align} \label{eq:freeaction}
	S_0 = \frac{1}{4\pi\alpha'} \int_\Sigma d^2 \sigma \sqrt{h} \, \Bigl( \CD X^{A'} \, \bar{\CD} X^{A'} + \lambda \, \bar{\CD} X + \bar{\lambda} \, \CD \overline{X} \Bigr) \,, 
\end{align}
where $\alpha'$ is the Regge slope. We defined $X \equiv X^0 + X^1$ and $\overline{X} \equiv X^0 - X^1$\,.
We also defined $\CD \equiv h^{-1/2} \, \epsilon^{\alpha\beta} \, \bigl( e_\alpha{}^\tau - i \, e_\alpha{}^\sigma \bigr) \, \p_\beta$ and $\bar{\CD} \equiv h^{-1/2} \, \epsilon^{\alpha\beta} \, \bigl( e_\alpha{}^\tau + i \, e_\alpha{}^\sigma \bigr) \, \p_\beta$\,.
Here, $e_\alpha{}^a$, $a = \tau, \sigma$ is the worldsheet Zweibein field that satisfies $h_{\alpha\beta} = e_\alpha{}^a \, e_\beta{}^a$\,. Moreover, $\epsilon^{\tau\sigma} = - \epsilon^{\sigma\tau} = 1$\,.
We will focus on the closed string sector in these notes and assume that the worldsheet $\Sigma$ is boundaryless.
For the theory to have a nonempty spectrum, the $X^1$-direction has to be compactified over a circle of radius $R$ \cite{Gomis:2000bd}. We assume that all the other directions remain uncompactified.

In conformal gauge, the free action \eqref{eq:freeaction} becomes \cite{Gomis:2000bd}
\be \label{eq:S0cg}
	S_0 = \frac{1}{4\pi\alpha'} \int_\Sigma d^2 \sigma \Bigl( \p X^{A'} \bar{\p} X^{A'} + \lambda \, \bar{\p} X + \bar{\lambda} \, \p \overline{X} \Bigr)\,,
\ee
with $\p \equiv i \, \p_\tau + \p_\sigma$ and $\bar{\p} \equiv - i \, \p_\tau + \p_\sigma$\,. %
The worldsheet $(1,0)$-form $\lambda$ and $(0,1)$-form $\bar{\lambda}$ play the role of a Lagrangian multiplier and impose the conditions 
\be \label{eq:chiralconds}
	\bar{\p} X = \p \overline{X} = 0\,, 
\ee
which are solved by $X = X(\tau + i \sigma)$ and $\overline{X} = \overline{X} (\tau - i \sigma)$\,.
These constraints are responsible for salient features of nonrelativistic string theory, including a string spectrum that enjoys a Galilean-invariant dispersion relation and intriguing localization theorems in the moduli space \cite{Gomis:2000bd}. 
Another direct consequence of these one-form fields is that the free theory \eqref{eq:S0cg} is invariant under an infinite number of spacetime isometries \cite{Batlle:2016iel}. We parametrize these spacetime isometries by (anti-)holomorphic functions $f(X)$\,, $\bar{f} (\overline{X})$\,, $g_{A'} (X)$\,, and $\bar{g}_{A'} (\overline{X})$\,. Supplemented with $\Lambda^{A'}{}_{\!B'}$ that parametrizes spatial rotations, the symmetry transformations acting on the worldsheet fields are
\begin{subequations} \label{eq:globaltrnsf}
\begin{align}
	\delta X^{A'} & = g^{A'} \! + \overline{g}^{A'} \! - \Lambda^{A'}{}_{\!B'} X^{B'},  \\[2pt]
	\delta X & = f\,,
		&
	& \hspace{-5cm} \delta \lambda = - \lambda \, \frac{df}{dX} - 2 \, \frac{dg^{}_{\!A'}}{dX} \, \p X^{A'}, \\[2pt]
	\delta \overline{X} & = \bar{f}\,,
		&
	& \hspace{-5cm} \delta \bar{\lambda} = - \bar{\lambda} \, \frac{d\bar{f}}{d\overline{X}} - 2 \,\frac{d\bar{g}^{}_{\!A'}}{d\overline{X}} \, \bar{\p} X^{A'}.
\end{align}
\end{subequations}
Note that there is no boost symmetry that transforms $X^A$ into $X^{A'}$, which implies that the full Lorentzian boost symmetry is absent. These transformations form the extended Galilean symmetry algebra that contains two copies of the Witt algebra \cite{Batlle:2016iel}. 

\subsection{One-form Fields and Dual Coordinates} \label{sec:ofdc}

A spacetime interpretation of the one-form worldsheet fields $\lambda$ and $\bar{\lambda}$ are made manifest by considering T-duality transformations in the longitudinal directions \cite{Bergshoeff:2018yvt}.  We already compactified $X^1$ over a circle. We now further compactify $X^0$ over a circle, which can be made sense of if one first perform a Wick rotation for $X^0$. The periodicity in the imaginary target-space time direction receives the physical interpretation as an inverse temperature.
To perform the T-dual of \eqref{eq:cg} along both longitudinal directions, we consider the parent action,
\begin{align} \label{eq:Spa0}
\begin{split}
	S_\text{parent} & = \frac{1}{4\pi\alpha'} \int_\Sigma d^2 \sigma \, \Bigl[ \p X^{A'} \bar{\p} X^{A'} + \lambda \, \bigl( \bar{u} + \bar{v} \bigr) + \bar{\lambda} \, \bigl( u - v \bigr) \Bigr] \\[2pt]
		& \quad + \frac{1}{2\pi\alpha'} \int_\Sigma d^2 \sigma \, \Bigl[ Y_0 \bigl( \bar{\p} u - \p \bar{u} \bigr) + Y_1 \bigl( \bar{\p} v - \p \bar{v} \bigr) \Bigr]\,. 
\end{split}
\end{align}
It is useful to introduce the following definitions:
\begin{align} \label{eq:gammarho}
	\gamma \equiv u - v\,,
		\qquad%
	\overline{\gamma} \equiv \bar{u} + \bar{v}\,;
		\qquad\qquad%
	\rho \equiv - u - v\,,
		\qquad%
	\bar{\rho} \equiv \bar{u} - \bar{v}\,.
\end{align}
In terms of the new variables in \eqref{eq:gammarho}, \eqref{eq:Spa0} becomes
\be \label{eq:hatS}
	S_\text{parent} = \frac{1}{4\pi\alpha'} \int_\Sigma d^2 \sigma \, \Bigl[ \p X^{A'} \bar{\p} X^{A'} + \lambda \, \bar{\gamma} + \bar{\lambda} \, \gamma - Y \bigl( \bar{\p} \rho + \p \bar{\gamma} \bigr) - \overline{Y} \bigl( \p \bar{\rho} - \bar{\p} \gamma \bigr) \Bigr]\,.
\ee
Integrating out $Y \equiv Y_0 + Y_1$ and $\overline{Y} \equiv Y_0 - Y_1$ in \eqref{eq:hatS} imposes $\bar{\p} \rho + \p \bar{\gamma} = \p \bar{\rho} - \bar{\p} \gamma = 0$\,, which can be solved locally by
\be \label{eq:rhogamma}
	\gamma = \p \overline{X}\,,
		\qquad
	\bar{\gamma} = \bar{\p} X\,;
		\qquad\qquad%
	\rho = - \p X\,,
		\qquad
	\bar{\rho} = \bar{\p} \overline{X}\,.
\ee
Plugging \eqref{eq:rhogamma} into \eqref{eq:hatS} recovers the original action \eqref{eq:S0cg}. Instead, we now integrate out the auxiliary fields $\gamma$ and $\bar{\gamma}$ in the parent action \eqref{eq:hatS}, which gives the dual action
\be \label{eq:Sdual}
	S_\text{dual} = \frac{1}{4\pi\alpha'} \int_\Sigma d^2 \sigma \Bigl( \p X^{A'} \bar{\p} X^{A'} + \rho \, \bar{\p} Y + \bar{\rho} \, \p \overline{Y} \Bigr)\,,
\ee
together with the relations,
\be \label{eq:lambdaYrel}
	\lambda = - \p Y\,, 
		\qquad
	\bar{\lambda} = \bar{\p} \overline{Y}\,.
\ee
The Lagrange multipliers $\rho$ and $\bar{\rho}$ in \eqref{eq:Sdual} impose the conditions $\bar{\p} Y = \p \overline{Y} = 0$\,.
The dual action \eqref{eq:Sdual} and also describes nonrelativistic string theory as the original action \eqref{eq:S0cg}, with the duality dictionary,
\be
	\lambda \longleftrightarrow \rho\,,
		\qquad
	\bar{\lambda} \longleftrightarrow \bar{\rho}\,,
		\qquad
	X \longleftrightarrow Y\,,
		\qquad
	\overline{X} \longleftrightarrow \overline{Y}\,.
\ee
The worldsheet fields $Y_A$ represent spacetime coordinates that are dual to $X^A$. This is consistent with the action \eqref{eq:ddaction} in Appendix \ref{app:Tdual}, where the duality transformation is preformed in two steps, first along $X^1$ and then $X^0$\,. There, a relation to the discrete light cone quantization (DLCQ) of relativistic string theory is made manifest \cite{Bergshoeff:2018yvt}. The relations in \eqref{eq:lambdaYrel} imply that $\lambda$ and $\bar{\lambda}$ encode the information of the dual worldsheet fields conjugate to string windings. 

In the original theory described by \eqref{eq:S0cg}, only $X^1$ is compactified over a circle of radius $R$\,.
Motivated by the T-duality transformations, but in the case where $X^0$ is not compactified, we introduce the field redefinitions,
\be \label{eq:lambdaY}
	\lambda = - \p Y\,,
		\qquad
	Y = Y (\tau+i\sigma)\,; 
		\qquad\qquad
	\bar{\lambda} = \bar{\p} \overline{Y}\,,
		\qquad
	\overline{Y} = \overline{Y}(\tau-i\sigma)\,.
\ee
These field redefinitions are in form the same as \eqref{eq:lambdaYrel}. We already learned that $Y$ and $\overline{Y}$ are conjugate to string windings while $X$ and $\overline{X}$ are conjugate to string momenta. See \cite{Gomis:2020fui, Gomis:2020izd} for related discussions in nonrelativistic open string theory.
 
The auxiliary coordinates $Y$ and $\overline{Y}$ are reminiscent of spacetime doublings in double field theory \cite{Hull:2009mi}, but only taking place in the longitudinal sector.  Nevertheless, na\"{i}vely plugging the field redefinitions \eqref{eq:lambdaY} back into the action \eqref{eq:S0cg} does \emph{not} lead to an equivalent theory. This is because the redefinitions \eqref{eq:lambdaY} involve time derivatives that would induce in the path integral a Jacobian $\det (\p \bar{\p})$\,, which contributes extra ghost terms in the action. Such ghosts compensate extra degrees of freedom introduced by the field redefinitions. 
In the following, we will keep using \eqref{eq:S0cg} as the defining action for nonrelativistic string theory, and we will always take the path integral to be defined with respect to $\lambda$ and $\bar{\lambda}$\,. The quantities $Y$ and $\overline{Y}$ are only introduced as auxiliary coordinates when winding modes are concerned.  

The same interpretation of $\lambda$ and $\bar{\lambda}$ can be extended to curved spacetime. The simplest way to incorporate the dual coordinates in the theory there is by performing a Hamiltonian analysis. Further details are included in Appendix \ref{app:Hfgm}. 

\subsection{Closed String Vertex Operators} \label{sec:OPE}

Interactions between strings are generated by turning on appropriate vertex operators. In order to consider strings propagating in a curved background, which is essentially a coherent state of strings, we need to classify first excited closed string vertex operators that are $(1,1)$-forms. Requiring that these vertex operators are BRST invariant derives the linearized spacetime equations of motion that dictate the target space dynamics. Before classifying the vertex operators, we first revisit the mode expansions \cite{Gomis:2000bd, Danielsson:2000mu} and introduce a set of worldsheet coordinates in which the operator product expansions (OPE) take a compact form. This formalism will facilitate later analysis of the OPEs.

\subsubsection*{Mode Expansions}

We start with revisiting the mode expansions of different worldsheet fields, including both $X^\mu$ and the dual coordinates $Y_A$\,. In radial quantization, we express \eqref{eq:S0cg} in terms of $z = e^{\tau + i \, \sigma}$ and $\bar{z} = e^{\tau - i \, \sigma}$, with
\be \label{eq:cg}
	S_0 = \frac{1}{4\pi\alpha'} \int_\mathbb{C} d^2 z \, \Bigl( 2 \, \p_z X^{A'} \, \p_{\bar{z}} X^{A'} + \lambda_z \, \p_{\bar{z}} X + \lambda_{\bar{z}} \, \p_z \overline{X} \Bigr) \,, 
\ee
where 
\be
	\lambda_z \equiv - \frac{i \lambda}{z} = - 2 \, \p_{z} Y\,,
		\qquad
	\lambda_{\bar{z}} \equiv \frac{i\bar{\lambda}}{\bar{z}} = 2 \, \p_{\bar{z}} \overline{Y}\,.
\ee
Since the $X^1$ direction is compactified, we have 
$X^1 (\sigma + 2\pi) = X^1 (\sigma) + 2 \pi R \, w$\,, 
$w \in \mathbb{Z}$\,.
The winding number $w$ is defined operatorially by
\begin{align} \label{eq:what}
	\hat{w} & = \frac{1}{2\pi R} \int_0^{2\pi} d\sigma \, \p_\sigma X^1 
	= \frac{1}{4\pi R} \oint_\CC \bigl( dz \, \p_z X - d\bar{z} \, \p_{\bar{z}} \overline{X} \bigr)\,,
\end{align}
where the contour $\CC$ is traversed counterclockwise along the string. Taking into account that $X = X(z)$ and $\overline{X} = \overline{X} (\bar{z})$\,, we find
\begin{subequations}
\begin{align}
	X (z) & = x + i \alpha' q \, \ln z + i \sqrt{2\alpha'} \sum_{m\neq0} \frac{\alpha_m}{m \, z^m}\,, \\[2pt]
	\overline{X} (\bar{z}) & = \overline{x} - i \alpha' \bar{q} \, \ln \bar{z} + i \sqrt{2\alpha'} \sum_{m\neq0} \frac{\tilde{\alpha}_m}{m \, \bar{z}^m}\,.
\end{align}
\end{subequations}
Since only the $X^1$ direction is compactified, we have $q = - \bar{q} = - wR / \alpha'$.
In general, $q$ and $-\bar{q}$ would take different values if $X^0$ were (Wick-rotated and) compactified.
The conjugate momenta for $X^A$ in \eqref{eq:cg} are defined operatorially via
\be
	{\hat{p}}^{}_0 = \frac{1}{4\pi\alpha'} \int_0^{2\pi} d\sigma \bigl( \lambda - \bar{\lambda} \bigr)\,,
		\qquad
	{\hat{p}}^{}_1 = \frac{1}{4\pi\alpha'} \int_0^{2\pi} d\sigma \bigl( \lambda + \bar{\lambda} \bigr)\,.
\ee
Recall that $Y = Y(z)$ and $\overline{Y} = \overline{Y} (\bar{z})$ are (anti-)holomorphic. Therefore,
\begin{subequations}
\begin{align}
	Y (z) & = y + i \alpha' p \, \ln z + i \sqrt{2\alpha'} \sum_{m\neq0} \frac{\beta_m}{m \, z^m}\,, \\[2pt]
	\overline{Y} (\bar{z}) & = \bar{y} - i \alpha' \bar{p} \, \ln \bar{z} + i \sqrt{2\alpha'} \sum_{m\neq0} \frac{\tilde{\beta}_m}{m \, {\bar{z}}^m}\,,
\end{align}
\end{subequations}
where $p = \frac{1}{2} \bigl( p^{}_0 + p^{}_1 \bigr)$ and $\bar{p} = \frac{1}{2} \bigl( p^{}_0 - p^{}_1 \bigr)$ are respectively eigenvalues of the operators 
\be \label{eq:pop}
	\hat{p} = - \frac{1}{2\pi\alpha'} \oint_\CC dz \, \p_z Y,
		\qquad
	\hat{\overline{p}} = - \frac{1}{2\pi\alpha'} \oint_\CC d\bar{z} \, \p_{\bar{z}} \overline{Y}.
\ee
In the compactified $X^1$ direction, the momentum eigenstates have $p_1 = n / R$\,, $n \in \mathbb{Z}$\,.

\subsubsection*{Operator Product Expansions}

The OPEs between $X^A$ and $Y_A$ are determined by the OPEs between $\lambda_z$\,, ${\lambda}_{\bar{z}}$\,, and $X^A$,
\be
	:\!\lambda_z (z_1) \, X(z_2)\!: \, \sim - \frac{2\alpha'}{z_{12}}\,,
		\qquad
	:\!\lambda_{\bar{z}} (\bar{z}_1) \, \overline{X}(\bar{z}_2)\!: \, \sim - \frac{2\alpha'}{\bar{z}_{12}}\,,
\ee
where $z_{ab} \equiv z_1 - z_2$ and $\bar{z}_{ab} \equiv \bar{z}_1 - \bar{z}_2$\,.
Using \eqref{eq:lambdaYrel}, we find the induced OPEs,
\be \label{eq:YXope}
	:\! Y (z_1) \, X(z_2) \!: \, \sim \alpha' \ln z_{12}\,,
		\qquad
	:\! \overline{Y} \, (\bar{z}_1) \, \overline{X}(\bar{z}_2) \!: \, \sim - \alpha' \ln \bar{z}_{12}\,.
\ee
It follows that, under a choice of branches,
\be
	[Y(z_1)\,, X(z_2)] = i \pi \alpha',
		\qquad
	[\overline{Y}(\bar{z}_1)\,, \overline{X}(\bar{z}_2)] = i \pi \alpha'.
\ee
Note that $Y_A$ and $X^A$ do not commute.
Moreover, we have
\be \label{eq:XXope}
	: \! X^{A'} \! (z_1\,, \bar{z}_1) \, X^{B'} \! (z_2\,, \bar{z}_2) \! : \, \sim - \tfrac{1}{2} \, \alpha' \, \delta^{A'B'} \ln |z_{12}|^2\,.
\ee
It is useful to write $X^{A'} = \co^{A'} (z) + \bar{\co}^{A'} (\bar{z})$ and define
\begin{subequations} \label{eq:XLR}
\begin{align} 
	\co^0 (z) & \equiv \tfrac{1}{2} (X + Y)\,, 
		&
	\overline{\co}^0 (\bar{z}) & \equiv \tfrac{1}{2} (\overline{X} - \overline{Y})\,, \\[2pt]
	\co^1 (z) & \equiv \tfrac{1}{2} (X - Y)\,,
		&
	\overline{\co}^1 (\bar{z}) & \equiv \tfrac{1}{2} (\overline{X} + \overline{Y})\,.
\end{align}
\end{subequations}
In terms of $\varphi^\mu$ and $\bar{\varphi}^\mu$, the OPEs in \eqref{eq:YXope} and \eqref{eq:XXope} become
\be
	:\!\co^\mu (z_1) \, \co^\nu (z_2)\!: \, \sim - \tfrac{1}{2} \, \alpha' \, \eta^{\mu\nu} \ln z_{12}\,,
		\qquad%
	:\!\overline{\co}^\mu (\bar{z}_1) \, \overline{\co}^\nu (\bar{z}_2)\!: \, \sim - \tfrac{1}{2} \, \alpha' \, \eta^{\mu\nu} \ln \bar{z}_{12}\,.
\ee
This formalism allows us to evaluate different OPEs in a compact way by directly borrowing the relativistic results. This would simplify the calculation in \cite{Gomis:2019zyu} and may also be useful for calculating amplitudes in matrix string theory \cite{saadlcq}.

\subsubsection*{Vertex Operators}

Finally, we study different closed string vertex operators, first the tachyon states and then the first excited states that give rise to background fields. We will also study the quantum consistency conditions required by BRST invariance. 

\vspace{3mm}

\noindent $\bullet$ \emph{Closed string tachyon states.} 
It is useful to group $Y_A$\,, $X^A$\,, and $X^{A'}$ into a single multiplet, $\mathbb{X}^{\text{I}} = (Y_A\,, X^A, X^{A'})^\intercal_{\phantom{\intercal}}$\,. In terms of this notation, the tachyonic operator is 
\begin{align} \label{eq:Vtach}
\begin{split}
	\CV & = e^{i \pi \, n \, \hat{w}} : \! e^{i \, \mathbb{K}^\text{I} \, \mathbb{X}^\text{I}} \!\!: \,,
\end{split}
\end{align}
where $\mathbb{K}^\text{I} \equiv \bigl( q^A,\, p^{}_A\,, \, k^{A'} \bigr)$\,, $q^0 \equiv q + \bar{q} = 0$\,, and $q^1 \equiv q - \bar{q} = -2wR/\alpha'$\,. For simplicity, we have set the coupling constant in \eqref{eq:Vtach} to one. This tachyonic operator corresponds to a common eigenstate for the operators \eqref{eq:what} and \eqref{eq:pop}. The phase factor $e^{i \pi w \hat{n}}$ is a cocycle factor that is required such that the vertex operators commute \cite{Polchinski:1998rq}. Since we will focus on the case where $w \neq 0$ but $n = 0$\,, the cocycle factor is set to 1. 

In terms of the change of variables in \eqref{eq:XLR}, we find
\be
	\CV = :\!\exp \bigl( i \, K^{}_{\mu} \, \co^\mu + i \, \overline{K}^{}_{\mu} \, \overline{\co}^\mu \bigr)\!:\,,
\ee
where $K_\mu = \bigl( \, p + q\,, \, p - q\,, k_{A'} \bigr)$ and $\overline{K}_\mu = \bigl( \, \bar{p} - \bar{q}\,, \, \bar{p} + \bar{q}\,, k_{A'} \bigr)$\,.
The BRST operator is
\be
	Q = \frac{1}{2\pi i} \oint \Bigl( dz \, \CJ - d \bar{z} \, \bar{\CJ} \Bigr)\,,
\ee
where $\CJ$ and $\bar{\CJ}$ are the BRST currents,
\begin{subequations}
\begin{align}
	\CJ & = c \, T^{\text{m}} + : \! b \, c \, \p_z c \! : + \tfrac{3}{2} \, \p_z^2 c\,,
		&%
	T^{\text{m}} & = - {\alpha'}{}^{-1} \, \eta_{\mu\nu} : \! \p_z \co^{\mu} \, \p_z \co^{\nu} \! :\,, \\[2pt]
	\bar{\CJ} & = \tilde{c} \, \tilde{T}^{\text{m}} + : \! \tilde{b} \, \tilde{c} \, \p_{\bar{z}} \tilde{c} \! : + \tfrac{3}{2} \, \p_{\bar{z}}^2 \tilde{c}\,,
		&%
	\tilde{T}^{\text{m}} & = - {\alpha'}{}^{-1} \, \eta_{\mu\nu} : \! \p_{\bar{z}} {\overline{\co}}^{\mu} \, \p_{\bar{z}} \overline{\co}^{\nu} \! :\,.
\end{align}
\end{subequations}
Here, $T^\text{m}$ and $\tilde{T}^\text{m}$ form the stress energy tensor. The Grassmannian fields $(b, c)$ and $(\tilde{b}, \tilde{c}\,)$ are the $bc$ ghosts. 
The BRST transformation of $\CV$ is given by
\be
	\bigl[ Q, \CV(z, \bar{z}) \bigr] = \p_z \bigl( c \, \CV \bigr) \! + \p_{\bar{z}} \bigl( \tilde{c} \, \CV \bigr) + \bigl( \mathcal{E} \, \p_z c + \tilde{\mathcal{E}} \, \p_{\bar{z}} \tilde{c} \, \bigr) \CV\,,
\ee 
where
\be
	\mathcal{E} = \tfrac{1}{4} \, \alpha' K^2 - 1\,,
		\qquad
	\tilde{\mathcal{E}} = \tfrac{1}{4} \, \alpha' \, {\overline{K}}{}^2 - 1\,,
\ee 
and $K^2 \equiv \eta_{\mu\nu} \, K^\mu \, K^\nu$ and ${\overline{K}}^2 \equiv \eta_{\mu\nu} \, {\overline{K}}^\mu \, {\overline{K}}^\nu$. BRST invariance requires that $[Q, \CV]$ be a total derivative, such that the integrated vertex operators remains unchanged. This imposes $\mathcal{E} = \tilde{\mathcal{E}} = 0$ and leads to the dispersion relation and level-matching condition,
\be \label{eq:nrdr}
	E = \frac{\alpha' \, k^2 - 4}{2 \, w R}\,,
		\qquad
	n \, w = 0\,,
\ee
where $k^2 \equiv k_{A'} k_{A'}$ and $E \equiv - p^{}_0$ denotes the energy.
For \eqref{eq:nrdr} to be well defined, we require $n = 0$ and $w \neq 0$\,, under which $K_\mu = \overline{K}_\mu = (\frac{p_0}{2} - \frac{w R}{\alpha'}\,, \frac{p_0}{2} + \frac{w R}{\alpha'}\,, k_{A'})$\,. This manifests its T-dual to the DLCQ of string theory. See more in Appendix \ref{app:Tdual}.

\vspace{3mm}

\noindent $\bullet$ \emph{First excited closed string states.} 
The corresponding vertex operators are
\be \label{eq:V1}
	\CV_{1} = g_{\mu\nu} \CV^{\mu\nu},
		\qquad
	\CV^{\mu\nu} = \, : \! \p_z \varphi^\mu \, \p_{\bar{z}} \overline{\varphi}^\nu \, \CV \!: \, .
\ee
The coefficient $g_{\mu\nu}$ can be decomposed into a symmetric tensor $s_{\mu\nu}$ and an anti-symmetric tensor $a_{\mu\nu}$\,, with $g_{\mu\nu} = s_{\mu\nu} + a_{\mu\nu}$\,. 
The BRST transformation of $\CV_{1}$ is
\begin{align} \label{eq:QV1comm}
\begin{split}
	\bigl[ Q, \CV_{1} (z, \bar{z}) \bigr] = \,\, & \p_z \Bigl( c \,\, \CV_{1} + i  \,  \CF_\mu \, \p_z c \,\, \p_{\bar{z}} \overline{\varphi}^\mu \, \CV \Bigr) + \mathcal{E}_{\mu\nu} \, \p_z c \, \CV^{\mu\nu} \\[2pt]	
	+ \, & \p_{\bar{z}} \Bigl( \tilde{c} \,\, \CV_{1} + i  \, \tilde{\CF}_\mu \, \p_{\bar{z}} \tilde{c} \,\, \p_{z} {\varphi}^\mu \, \CV \Bigr) + \tilde{\mathcal{E}}_{\mu\nu} \, \p_{\bar{z}} \tilde{c} \,\, \CV^{\mu\nu},
\end{split}
\end{align}
where
\begin{subequations}
\begin{align}
	\CF_\mu & = - \tfrac{1}{4} \, \alpha' \, K^\rho \, g_{\rho\mu}\,, 
		&%
	\mathcal{E}_{\mu\nu} & = \tfrac{1}{4} \, \alpha' \bigl( K^2 \, g_{\mu\nu} - K_\mu \, K^\rho \, g_{\rho\nu} \bigr) \,, \\[2pt]
	\tilde{\CF}_\mu & = - \tfrac{1}{4} \, \alpha' \, {\overline{K}}^\rho \, g_{\mu\rho}\,, 
		&%
	\tilde{\mathcal{E}}_{\mu\nu} & = \tfrac{1}{4} \, \alpha' \bigl( {\overline{K}}^2 \, g_{\mu\nu} - \overline{K}_\nu \, \overline{K}^\rho \, g_{\mu\rho} \bigr)\,.
\end{align}
\end{subequations}
BRST invariance requires $\mathcal{E}_{\mu\nu} = \tilde{\mathcal{E}}_{\mu\nu} = 0$\,, which are the linearized equations of motion in Fourier space.
Moreover, the vertex operator $\CV_{1}$ gains a quantum correction at order $\alpha'$. Requiring that the vertex operator remains unchanged imposes the gauge-fixing conditions $\CF_\mu = \tilde{\CF}_\mu = 0$\,. We thus find that the corresponding physical states satisfy
\be
	K^\rho \, g_{\rho\mu} = \overline{K}^\rho \, g_{\mu\rho} = 0\,,
		\qquad
	E = \frac{\alpha' k^2}{2 w R}\,,
		\qquad
	n w = 0\,.
\ee
For the dispersion relation to be well defined, we require $w \neq 0$\,, which implies $n = 0$\,. In general, all components of \eqref{eq:V1} have to be included in the spectrum, such that the vertex operators are closed under OPEs.

\subsection{Deformation Towards Relativistic String Theory} \label{sec:dtrst}

In addition to the vertex operators that represent physical asymptotic states with nonzero windings in the $X^1$ direction, there is also a zero-winding sector that contains intermediate states carrying instantaneous Newtonian-like interactions. These zero-winding states are of measure zero in the asymptotic limit and cannot be put on-shell \cite{Gomis:2000bd, Danielsson:2000mu}. Inserting the vertex operators \eqref{eq:V1} with a zero winding number in the path integral leads to sigma models in geometry and Kalb-Ramond background fields \cite{Gomis:2019zyu}. The dynamics of such background fields define the EFTs in spacetime, which contain no propagating degrees of freedom. In the sigma model, these background fields contribute quadratic terms that modify the free action \eqref{eq:cg}. For example, one may turn on the Lagrangian terms
$\p_z \overline{X} \, \p_{\bar{z}} {X}$ and $\p_z X \, \p_{\bar{z}} \overline{X}$\,. The difference between these two terms is $2 \, \p_z X^A \, \p_{\bar{z}} X^B \epsilon_{AB}$\,,
which corresponds to a constant $B$-field in nonrelativistic string theory. We defined $\epsilon_{01} = - \epsilon_{10} = 1$\,. Therefore, we only need to worry about whether adding the Lagrangian term $\p_z X \, \p_{\bar{z}} \overline{X}$ to \eqref{eq:cg} changes the nature of the theory. The deformed action is
\be \label{eq:S'0}
	S'_0 = \frac{1}{4\pi\alpha'} \int_\mathbb{C} d^2 z \Bigl( \p_\alpha X^{A'} \p^\alpha X^{A'} + \lambda_z \, \p_{\bar{z}} X + \lambda_{\bar{z}} \, \p_z \overline{X} + \eta \, \p_z \overline{X} \, \p_{\bar{z}} {X} \Bigr)\,,
\ee
where $\p_z \overline{X} \, \p_{\bar{z}} {X}$ can be removed by performing a field redefinition, $\lambda_z \rightarrow \lambda_z - \eta \, \p_z \overline{X}$\,. Hence, $S'_0$ in \eqref{eq:S'0} is equivalent to \eqref{eq:cg} that defines nonrelativistic string theory. 

There exists another deformation that does change the nature of the theory. Note that the general vertex operator \eqref{eq:V1} contains a term in the zero-winding sector
\be
	- \tfrac{1}{4} \lr s_{00} - 2 \, s_{01} + s_{11} \rr \, : \! \p_z Y \, \p_{\bar{z}} \overline{Y} \exp \bigl( i \, p^{}_{\!A} X^A + i \, k_{\!A'} X^{A'} \bigr) \! :\,, 
\ee
which contributes the marginal interacting term
\be
	S_{\lambda\bar{\lambda}} = \frac{1}{8\pi\alpha'} \int_\mathbb{C} d^2 z \, \lambda_z \, \lambda_{\bar{z}} \, U[X]
\ee
to the sigma model. We already summed over all momentum states.
We have replaced $\p_\alpha Y^A$ with $\lambda_z$ and $\lambda_{\bar{z}}$\,, with respect to which the path integral is defined. However, when the coupling $U[X] = U_0$ is constant, $S_{\lambda\bar{\lambda}}$ is a quadratic term and already modifies the free theory in \eqref{eq:cg}. The deformed free action is
\be \label{eq:defaction}
	S_\text{def.} = \frac{1}{4\pi\alpha'} \int_\mathbb{C} d^2 z \, \lr 2 \, \p_z X^{A'} \, \p_{\bar{z}} X^{A'} + \lambda_z \, \p_{\bar{z}} X + \lambda_{\bar{z}} \, \p_z \overline{X} + \tfrac{1}{2} \, U_0 \, \lambda_z \, \lambda_{\bar{z}} \rr.
\ee
The dispersion relation in the string spectrum now receives a $U_0$ deformation \cite{Grignani:2001hb}, 
\be
	E = \frac{\alpha'}{2 wR} \ls k^2 + \frac{2}{\alpha'} \, \bigl( N + \tilde{N} - 2 \bigr) - U_0 \, \lr E^2 - \frac{n^2}{R^2} \rr \rs,
\ee
where we reintroduced a nonzero winding number. Here, $N$ and $\tilde{N}$ denote the string excitation numbers. In contrast to nonrelativistic string theory that arises at $U_0 = 0$\,, where all physical states carry a nonzero winding number, now, there are also asymptotic states with $w = 0$\,. These zero-winding states enjoy a well-defined relativistic dispersion relation (we require that $U_0 > 0$ for stability), 
\be \label{eq:reldr}
	U_0 \, E^2 - k^2 = U_0 \, \frac{n^2}{R^2} + \frac{2}{\alpha'} \bigl( N + \tilde{N} - 2 \bigr)\,.
\ee 
In this sense, a nonzero $U_0$ deforms the theory towards the relativistic regime. 

In the free theory, one can always tune $U_0$ to zero. However, the $\lambda\bar{\lambda}$ operator will be generated via OPEs when interactions are turned on by inserting various vertex operators on the worldsheet, unless additional symmetry principles are applied. For example, consider $\CV_1 = s^{}_{A'0} \, \p_z \varphi^{A'} \p_{\bar{z}} \bar{\varphi}^0$\,, which does not contain $\p_z Y \, \p_{\bar{z}} \overline{Y}$\,; however, the commutation relation $[Q, \CV_1]$ in \eqref{eq:QV1comm} generates $- \frac{w R}{8} \, k^{}_{A'} \, s^{}_{A'0} \, \p_z Y \, \p_{\bar{z}} \overline{Y}$\,. This quantum contribution that is linear in $s^{}_{\mu\nu}$ vanishes when $w = 0$\,, but we will see later in \S\ref{sec:nscb} that there are nonzero quantum contributions of higher orders in the background field fluctuations even in the zero-winding sector. Therefore, the $\lambda\bar{\lambda}$ operator needs to be included in the theory unless additional symmetries are imposed.
This implies that in general $U_0 \neq 0$ in \eqref{eq:defaction}, in which case the one-form fields $\lambda$ and $\bar{\lambda}$ remain (anti-)holomorphic, while the conditions \eqref{eq:chiralconds} are deformed to be
\be
	\p_{\bar{z}} X = - \tfrac{1}{2} \, U_0 \, \lambda_{\bar{z}}\,,
		\qquad
	\p_z \overline{X} = - \tfrac{1}{2} \, U_0 \, \lambda_z\,,
\ee
and all the analysis performed earlier in this section will be have to be modified accordingly. The deformed action \eqref{eq:defaction} enjoys the global symmetries,
\begin{subequations}
\begin{align}
	\delta X^A & = \Theta^A + \Lambda \, \epsilon^A{}_B \, X^B + U_0 \, \Lambda^A{}_{A'} \, X^{A'}\!, 
		&
	\delta \lambda_z & = \Lambda \, \lambda_z + 2 \, {\Lambda}^{A'} \, \p_z X^{A'}\!, \\[2pt]
	\delta X^{A'} & = \Theta^{A'} + \Lambda^{A'}{}_A \, X^A + \Lambda^{A'}{}_{\!B'} X^{B'}\!, 
		&
	\delta \lambda_{\bar{z}} & = -  \Lambda \, \lambda_{\bar{z}} + 2 \, \bar{\Lambda}^{A'} \, \p_{\bar{z}} X^{A'}\!, 
\end{align}
\end{subequations}
where the full spacetime Lorentz boost transformation arises. Here, $\Lambda_{A'} \equiv \Lambda_0{}^{A'} + \Lambda_1{}^{A'}$ and $\bar{\Lambda}_{A'} \equiv \Lambda_0{}^{A'} - \Lambda_1{}^{A'}$ parametrize the boost transformations between the longitudinal and transverse sectors, $\Lambda$ parametrizes the longitudinal Lorentz transformation, $\Lambda^{A'}{}_{\,B'}$ parametrizes the transverse rotations, and $\Theta^A$ and $\Theta^{A'}$ parametrize the longitudinal and transverse translations, respectively. The underlying symmetry is given by the Poincar\'{e} algebra, which can not be embedded in the infinite-dimensional algebra associated with the transformations in \eqref{eq:globaltrnsf}, unless the contraction $U_0 = 0$ is applied. The relevant commutation relations between different generators are given in Appendix \ref{app:pa}{\color{blue}.1}. It is also interesting to note that the transformations of $\lambda$ and $\bar{\lambda}$ induce
\be
	\delta Y_A = \tilde{\Theta}_A + \epsilon_A{}^B \bigl( \Lambda \, Y_B - \Lambda_{BA'} X^{A'} \bigr)\,,
\ee
where a boost that transforms $Y_A$ into $X^{A'}$ (but not \emph{vice versa}) emerges.
Finally, integrating out $\lambda_z$ and $\lambda_{\bar{z}}$ in \eqref{eq:defaction} leads to the equivalent action
\be
	S_\text{def.} = \frac{1}{2\pi\alpha'} \int_\mathbb{C} d^2 z \, \Bigl[ \p_z X^{A'} \, \p_{\bar{z}} X^{A'} + U_0^{-1} \lr \eta_{AB} - \epsilon_{AB} \rr \p_z X^A \, \p_{\bar{z}} X^B \Bigr]\,,
\ee
which is manifestly relativistic string theory in a constant $B$-field. 

As we have seen above, the $\lambda\bar{\lambda}$ term deforms nonrelativistic string theory towards relativistic string theory. If one wishes to define nonrelativistic string theory as a self-contained corner of relativistic string theory by a renormalizable worldsheet QFT, a symmetry argument is required to protect the sigma model from the $\lambda\bar{\lambda}$ deformation at both the classical and quantum level. We will discuss this in detail in \S\ref{sec:nscb}, where we will restrict ourselves to the zero-winding sector and consider interacting two-dimensional QFTs that perturb around the free fixed point defined by \eqref{eq:freeaction}. We will see that the $\lambda\bar{\lambda}$ term is generated by log-divergent quantum corrections that correspond to spacetime torsions that twist the foliation structure in the target space \cite{Gomis:2019zyu, Yan:2019xsf, Gallegos:2019icg}. We will therefore refer to such a deformation proportional to $\lambda \bar{\lambda}$ as the \emph{torsional deformation}. 

\section{Nonrelativistic Strings in Curved Backgrounds} \label{sec:nscb}

After reviewing how the torsional deformation $\lambda\bar{\lambda}$ deforms nonrelativistic string theory towards relativistic string theory, we now consider strings in arbitrary geometry, $B$-field, and dilaton backgrounds, focusing on the zero-winding sector. We will consider target space gauge symmetries under which a renormalizable worldsheet QFT without the $\lambda \bar{\lambda}$ term is defined. We start with sigma models in unconstrained backgrounds that incorporate the $\lambda\bar{\lambda}$ operator, associated with a functional coupling $U = U[X]$\,. These sigma models describe relativistic string theory but in an unconventional parametrization, which is however useful for accessing the corner of nonrelativistic strings at $U = 0$\,. We then analyze the RG properties in the limit $U \rightarrow 0$\,. This perspective rooted in relativistic string theory will provide us with intuition for constructing the spacetime gauge symmetries for defining nonrelativistic string theory from first principles. 

\subsection{String Sigma Models in General Background Fields} \label{sec:ssm}

Turning on interactions in the free action \eqref{eq:S0cg} gives rise to background fields to which nonrelativistic string theory is coupled. Allowing the most general marginal terms that are compatible with the worldsheet diffeomorphisms and (classical) conformal symmetry in the sigma model, we obtain
\begin{align} \label{eq:Sint}
\begin{split}
	& S = \frac{1}{4\pi\alpha'} \int_\Sigma d^2 \sigma \, \sqrt{h} \, \CD X^\mu \, \bar{\CD} X^\nu \bigl( S_{\mu\nu}[X] + A_{\mu\nu}[X] \bigr) \\[2pt]
	& \quad\! + \frac{1}{4\pi\alpha'} \int_\Sigma d^2 \sigma \, \sqrt{h} \, \Bigl\{ \lambda \, \bar{\CD} X^\mu \, \tau_\mu[X] + \bar{\lambda} \, \CD X^\mu \, \bar{\tau}_\mu[X] + \lambda \, \bar{\lambda} \, U[X] + \alpha' \, \text{R}^{(2)} \, \Phi[X] \Bigr\}\,,
\end{split}
\end{align}
where $S_{\mu\nu}$ is symmetric and $A_{\mu\nu}$ is antisymmetric. We defined R${}^{(2)}$ as the Ricci scalar associated with the worldsheet metric $h_{\alpha\beta}$\,. The action \eqref{eq:Sint} describes the full relativistic string theory but with an unconventional parametrization of background fields. 
 We have required the QFT to be local, which means that the background fields are functionals of $X^\mu$ but not the dual coordinates $Y_A$ (with its incarnation in curved spacetime).~\footnote{\,Note that $Y_A$ are integrals of the local fields $\lambda$ and $\bar{\lambda}$\,. In this sense, $Y_A$ are nonlocal.} More generally, nonlocal dependences on $\lambda$ and $\bar{\lambda}$ can also be introduced in the backgrounds, by exponentiating the vertex operators in \eqref{eq:V1}. This will lead us to generalized geometry where the dual coordinates $Y$ and $\overline{Y}$ become visible. We will briefly discuss the connection to generalized metric in Appendix \ref{app:Hfgm}. 

Define $\tau_\mu{}^0 = \tfrac{1}{2} \bigl( \tau_\mu + {\bar{\tau}}_\mu \bigr)$ and $\tau_\mu{}^1 = \tfrac{1}{2} \bigl( \tau_\mu - {\bar{\tau}}_\mu \bigr)$\,. The functional coupling $\tau_\mu{}^A$ can be thought of as a longitudinal Vielbein field. Under reparametrizations of $X^\mu$\,, we have
\begin{subequations}
\begin{align}
	S_{\mu\nu} [\tilde{X}] & = \frac{\p X^\rho}{\p \tilde{X}^\mu} \frac{\p X^\sigma}{\p \tilde{X}^\nu} \, S_{\rho\sigma} [X]\,,
		&%
	\tau_\mu{}^A [\tilde{X}] & = \frac{\p X^\rho}{\p \tilde{X}^\mu} \, \tau_{\rho}{}^A [X]\,, \\[2pt]
	A_{\mu\nu} [\tilde{X}] & = \frac{\p X^\rho}{\p \tilde{X}^\mu} \frac{\p X^\sigma}{\p \tilde{X}^\nu} \, A_{\rho\sigma} [X]\,,
		&%
	U[\tilde{X}] & = U[X]\,, 
		&%
	\Phi [\tilde{X}] & = \Phi [X]\,.
\end{align}
\end{subequations}
The parametrization of the background fields in $S_{\mu\nu}$\,, $A_{\mu\nu}$\,, $\tau_\mu{}^A$\,, $U$, and $\Phi$ is redundant. This is because $\lambda$ and $\bar{\lambda}$ are defined up to (finite) field redefinitions, which we parametrize by $\CC$\,, $\overline{\CC}$\,, $\CC_\mu = \CC_\mu{}^0 + \CC_\mu{}^1$, and $\overline{\CC}_\mu = \CC_\mu{}^0 - \CC_\mu{}^1$,
\begin{subequations} \label{eq:stueckelberg}
\be 
	\lambda \rightarrow \CC^{-1} \bigl( \lambda - \CD X^\mu \, \bar{\CC}_\mu \bigr)\,,
		\qquad
	\bar{\lambda} \rightarrow \bar{\CC}^{-1} \bigl( {\bar{\lambda}} - \bar{\CD} X^\mu \, \CC_\mu \bigr)\,.
\ee
Simultaneously, we take the redefinitions of the background fields as follows:
\begin{align}
	S_{\mu\nu} & \rightarrow S_{\mu\nu} - \bigl( \CC_\mu{}^A \, \tau_\nu{}^B \! + \CC_\nu{}^A \, \tau_\mu{}^B + \CC_\mu{}^A \, \CC_\nu{}^B \, U \bigr) \, \eta^{}_{AB} \,, 
		&%
	\tau^{}_\mu & \rightarrow \CC \, \bigl( \tau_\mu + \CC_\mu \, U \bigr)\,, \\[2pt]
	A_{\mu\nu} & \rightarrow A_{\mu\nu} + \bigl( \CC_\mu{}^A \, \tau_\nu{}^B \! - \CC_\nu{}^A \, \tau_\mu{}^B +  \CC_\mu{}^A \, \CC_\nu{}^B \, U \bigr) \, \epsilon^{}_{AB} \,,
		&%
	\bar{\tau}^{}_\mu & \rightarrow \bar{\CC} \, \bigl( \bar{\tau}_\mu + \bar{\CC}_\mu \, U \bigr)\,, 
\end{align}
and
\begin{align}
	U & \rightarrow \CC \, \bar{\CC} \, U\,,
		&
	\Phi & \rightarrow \Phi + \tfrac{1}{2} \ln |\CC \bar{\CC}|\,.
\end{align}
\end{subequations}
Here, $\CC / \bar{\CC}$ and $\CC \bar{\CC}$ parametrize a Lorentz boost in the longitudinal sector and a dilatational transformation, respectively. 
Under the above redefinitions that can be treated as Stueckelberg symmetries, the action \eqref{eq:Sint} remains unchanged.  

To further the discussion, it is useful to introduce a transverse Vielbein field $E_\mu{}^{A'}$ that is orthogonal to $\tau_\mu{}^A$\,, in the sense that, together with the inverse Vielbein fields $E^\mu{}^{}_{\!A'}$ and $\tau^\mu{}_A$\,, we have the following invertibility conditions:
\begin{subequations} 
\begin{align}
	\tau^\mu{}^{}_{\!A} \, \tau_\mu{}^B & = \delta^B_A\,, 
		&%
	\tau_\mu{}^A \, \tau^\nu{}^{}_{\!\!A} + E_\mu{}^{A'} \, E^\nu{}^{}_{\!\!A'} & = \delta_\mu^\nu\,, \\[2pt]
	E^\mu{}^{}_{\!A'} \, E^{}_\mu{}^{\!B'} & = \delta_{A'}^{B'},
		&%
	\tau^\mu{}^{}_{\!A} \, E^{}_\mu{}^{A'} = E^\mu{}^{}_{\!A'} \, \tau^{}_\mu{}^A & = 0\,.
\end{align}
\end{subequations}
We choose $\CC_{\mu}{}^A$ in \eqref{eq:stueckelberg} such that the transformed $S_{\mu\nu} \, \tau^\nu{}_{\!A}$ vanishes\,. We therefore write the string action with the fixed $\CC_\mu{}^A$ Stueckelberg symmetry as \cite{Yan:2019xsf, Bergshoeff:2021bmc}
\begin{align}  \label{eq:SEA}
\begin{split}
	S & = \frac{1}{4\pi\alpha'} \int_\Sigma d^2 \sigma \, \sqrt{h} \, \CD X^\mu \, \bar{\CD} X^\nu \Bigl( E_{\mu\nu} + \CA_{\mu\nu} \Bigr) \\[2pt]
	& \quad + \frac{1}{4\pi\alpha'} \int_\Sigma d^2 \sigma \, \sqrt{h} \, \Bigl( \lambda \, \bar{\CD} X^\mu \, \tau_\mu + \bar{\lambda} \, \CD X^\mu \, \bar{\tau}_\mu + \lambda \, \bar{\lambda} \, U + \alpha' \, \text{R}^{(2)} \, \Phi \Bigr)\,,
\end{split}
\end{align}
where $E_{\mu\nu}$ is symmetric and $\CA_{\mu\nu}$ is anti-symmetric. Here, $E_{\mu\nu}$ satisfies
$E_{\mu\nu} \, \tau^\nu{}_{\!A} = 0$\,. 
We therefore choose the reparametrization of $X^\mu$ such that $E_{\mu\nu} = E_\mu{}^{A'} E_\nu{}^{A'}$.
It is also possible to further apply the dilatational symmetry parametrized by $\CC \bar{\CC}$ such that $U = 1$\,, in which case the corner that describes nonrelativistic strings at the limit $U \rightarrow 0$ is invisible. Bearing in mind that at the end we aim to take the $U \rightarrow 0$ limit in this formalism, we will leave this dilatational symmetry unfixed. 

Integrating out $\lambda$ and $\bar{\lambda}$ in \eqref{eq:Sint} gives the standard string action,
\begin{align} \label{eq:Srel}
	S & = \frac{1}{4\pi\alpha'} \int_\Sigma d^2 \sigma \, \sqrt{h} \, \Bigl[ \CD X^\mu \, \bar{\CD} X^\nu \bigl( \hat{G}_{\mu\nu} + \hat{B}_{\mu\nu} \bigr) + \alpha' R \, \hat{\Phi} \Bigr]\,, 
\end{align}
where
\be \label{eq:GBP}
	\hat{G}_{\mu\nu} = \frac{1}{U} \, \tau_{\mu\nu} + E_{\mu\nu}\,,
		\qquad
	\hat{B}_{\mu\nu} = - \frac{1}{U} \, \tau_\mu{}^A \, \tau_\nu{}^B \epsilon_{AB} + \CA_{\mu\nu}\,,
		\qquad
	\hat{\Phi} = \Phi - \frac{1}{2} \ln |U|\,.
\ee
In \eqref{eq:Srel}, it is not manifest that $U \rightarrow 0$ is nonsingular. In contrast, the equivalent formalism \eqref{eq:SEA} of relativistic string theory gives a natural framework for zooming in the regime around $U = 0$\,.  
To study the spacetime gauge symmetries, it is convenient to introduce a Vielbein field $\hat{E}_\mu{}^{M}$ that satisfies $\hat{G}_{\mu\nu} = \hat{E}_\mu{}^{\hat{A}} \hat{E}_\nu{}^{\hat{B}} \, \eta_{\hat{A}\hat{B}}$, with $\hat{A} = (A, A')$\,. 
The spacetime local gauge transformations of the background fields in \eqref{eq:SEA} are then induced by the local Lorentz transformation 
$\delta_{\hat{\Lambda}} \hat{E}_\mu{}^{\hat{A}} = \hat{\Lambda}^{\hat{A}}{}_{\hat{B}} \, \hat{E}_\mu{}^{\hat{B}}$,
and the $U(1)$ two-form transformation $\delta_{\epsilon} \hat{B}_{\mu\nu} = \p_\mu \epsilon_\nu - \p_\nu \epsilon_\mu$\,.  
It is understood that all the quantities are covariant under the diffeomorphisms, which we will therefore omit in the following discussions.  
Without loss of generalities, we set $U > 0$\,. Identifying
\begin{subequations}
\begin{align}
	\hat{E}_\mu{}^A & = U^{-1/2} \, \tau_\mu{}^A\,, 
		&%
	\hat{\Lambda}^A{}_B & = \Lambda^A{}_B\,, \\[2pt]
	\hat{E}_\mu{}^{A'} & = E_\mu{}^{A'},
		&%
	\hat{\Lambda}^A{}_{A'} & = U^{1/2} \, \Lambda^A{}_{A'}\,, 
		&%
	\hat{\Lambda}^{A'}{}_{B'} & = \Lambda^{A'}{}_{B'}\,,
\end{align}
\end{subequations}
we find the induced infinitesimal spacetime gauge transformations,
\begin{subequations} \label{eq:deltasnc}
\begin{align}
	\delta \tau_\mu{}^A & = \Lambda_\text{D} \, \tau_\mu{}^A + \Lambda \, \epsilon^A{}_B \, \tau_\mu{}^B + U \, \Lambda^A{}_{\!A'} \, E_\mu{}^{A'}, 
		&%
	\delta U & = 2 \, \Lambda_\text{D} \, U\,, \\[2pt]
	\delta E_\mu{}^{A'} & = \Lambda^{A'}{}_{\!A} \, \tau_\mu{}^A + \Lambda^{A'}{}_{\!B'} \, E_\mu{}^{B'}, 
		&%
	\delta \Phi & = \Lambda_\text{D} \, \Phi\,, \label{eq:deltasncPhi} \\[2pt]
	\delta \CA_{\mu\nu} & = \Lambda^A{}_{A'} \, \epsilon^{}_{AB} \, \bigl( E_\mu{}^{A'} \, \tau_\nu{}^B - E_\nu{}^{A'} \, \tau_\mu{}^B \bigr)\,.
\end{align}
\end{subequations}
Here, $\Lambda \equiv - \tfrac{1}{2} \, \epsilon_A{}^B \Lambda^A{}_B$\,. Note that we also have an emergent dilatational symmetry parametrized by $\Lambda_\text{D}$\,.
For the action \eqref{eq:SEA} to be invariant under \eqref{eq:deltasnc}, we require
\begin{subequations}
\begin{align}
	\delta \lambda & = - \bigl( \Lambda_\text{D} - \Lambda \bigr) \lambda + \Lambda_{A'} \, E_\mu{}^{A'} \CD X^\mu, \\[2pt]
	\delta \bar{\lambda} & = - \bigl( \Lambda_\text{D} + \Lambda \bigr) \bar{\lambda} + \bar{\Lambda}_{A'} \, E_\mu{}^{A'} \bar{\CD} X^\mu.
\end{align}
\end{subequations}
It is useful to introduce an additional gauge field $m_\mu{}^A$ that transforms as
\be \label{eq:mmuA}
	\delta m_\mu{}^A = - \Lambda_\text{D} \, m_\mu{}^A + \Lambda \, \epsilon^A{}_B \, m_\mu{}^B + \Lambda^A{}_{\!A'} \, E_\mu{}^{A'},
\ee
such that
\be
	B_{\mu\nu} \equiv \CA_{\mu\nu} - \bigl( m_\mu{}^A \, \tau_\nu{}^B - m_\nu{}^A \, \tau_\mu{}^B - U \, m_\mu{}^A \, m_\nu{}^B \bigr) \, \epsilon_{AB}
\ee
is invariant under $\Lambda_\text{D}$\,, $\Lambda$\,, $\Lambda^A{}_{A'}$\,, and $\Lambda^{A'}{}_{B'}$\,. This new field $B_{\mu\nu}$ only transforms under diffeomorphisms and the $U(1)$ two-form symmetry as $\delta_{\epsilon} B_{\mu\nu} = \p_\mu \epsilon_\nu - \p_\nu \epsilon_\mu$\,, and will be identified as the Kalb-Ramond field in nonrelativistic string theory.\,\footnote{\,In \cite{Bergshoeff:2021bmc}, the background field $\CA_{\mu\nu}$ (instead of $B_{\mu\nu}$) in these notes is referred to as the Kalb-Ramond field, which transforms nontrivially under local spacetime gauge symmetries. In fact, when $U=0$\,, the geometry background and Kalb-Ramond field are intertwined with each other \cite{Bergshoeff:2021bmc}. How one splits the background fields between the geometrical data and the Kalb-Ramond field in the $U \rightarrow 0$ limit is a matter of choice. We refer to $B_{\mu\nu}$ as the Kalb-Ramond field in these notes so that it is cleaner to present the gauge transformations of various background fields.} The underlying symmetry algebra is the Poincar\'{e} group augmented with a longitudinal dilatational symmetry, which is defined in Appendix \ref{app:pa}{\color{blue}.1}. 

\subsection{Beta-Functionals and Torsional Deformations} \label{sec:bftd}

We now move on to the quantum aspects of the action \eqref{eq:SEA} and study the renormalization group flow structure. Not all components of the background fields in the two-dimensional QFT defined by \eqref{eq:SEA} have independent beta-functionals. This is due to the path-integral identities,
\begin{subequations}
\begin{align}
	0 & = \int \frac{\delta}{\delta \lambda} \lr e^{-S} \, \CD X^\mu \rr = \int \frac{\delta}{\delta \lambda} \lr e^{-S} \, \lambda \rr, \\[2pt]
	0 & = \int \frac{\delta}{\delta \bar{\lambda}} \lr e^{-S} \, \bar{\CD} X^\mu \rr = \int \frac{\delta}{\delta \bar{\lambda}} \lr e^{-S} \, \bar{\lambda} \rr.
\end{align}
\end{subequations}
It follows that
\begin{subequations} \label{eq:id1}
\begin{align} 
	\langle \bar{\CD} X^\mu \, \CD X^\nu \, \tau_\nu \, \cdots \, \rangle & = - \langle U \, \lambda \, \bar{\CD} X^\mu \cdots \rangle\,, \\[2pt]
	\langle \CD X^\mu \, \bar{\CD} X^\nu \, \tau_\nu \cdots \rangle & = - \langle U \, \bar{\lambda} \, \CD X^\mu \cdots \rangle\,,
\end{align}
\end{subequations}
and
\begin{subequations}  \label{eq:ll}
\begin{align}
	\langle \lambda \, \bar{\CD} X^\mu \, \tau_\mu \cdots \rangle + \langle \lambda \bar{\lambda} \, U  \cdots \rangle & = \langle \delta^{(2)} (0) \cdots \rangle\,, \\[2pt]
	\langle \bar{\lambda} \, \CD X^\mu \, \overline{\tau}_\mu \cdots \rangle + \langle \lambda \bar{\lambda} \, U  \cdots \rangle & = \langle \delta^{(2)} (0) \cdots \rangle\,.
\end{align}
\end{subequations}
Here, ``$\cdots$" denotes insertions of other operators.
From \eqref{eq:ll} we find 
\be \label{eq:id2}
	\langle \lambda \, \bar{\CD} X^\mu \, \tau_\mu \cdots \rangle = \langle \bar{\lambda} \, \CD X^\mu \, \overline{\tau}_\mu \cdots \rangle\,.
\ee
These path-integral identities are analogous to the ones considered in \cite{Gomis:2019zyu, Yan:2019xsf}. A rigorous treatment of these identities can be found in \cite{Yan:2019xsf}, only with slight modifications.
The origin of these path-integral identities can be attributed to the Stueckelberg symmetry in \eqref{eq:stueckelberg}. The one-loop effective action is
\begin{align}  \label{eq:olefa}
\begin{split}
	\Gamma_\text{1-loop} & = - \frac{\log\Lambda}{4\pi\alpha'} \int_\Sigma d^2 \sigma \, \sqrt{h} \, \CD X^\mu \, \bar{\CD} X^\nu \Bigl( \beta^E_{\mu\nu} + \beta^\CA_{\mu\nu} \Bigr) \\[2pt]
	& \quad - \frac{\log \Lambda}{4\pi\alpha'} \int_\Sigma d^2 \sigma \, \sqrt{h} \, \Bigl( \lambda \, \bar{\CD} X^\mu \, \beta^\tau_\mu + \bar{\lambda} \, \CD X^\mu \, \beta^{\bar{\tau}}_\mu + \lambda \, \bar{\lambda} \, \beta^U + \alpha' \, \text{R}^{(2)} \, \beta^F \Bigr)\,.
\end{split}
\end{align}
Here, $\Lambda$ denotes the UV cutoff scale. Define
\be \label{eq:defF}
	F = \Phi - \tfrac{1}{4} \ln G\,,
		\qquad
	G = \ls\det \begin{pmatrix} 
		\tau_\mu{}^A \\[2pt]
		E_\mu{}^{A'} 
		\end{pmatrix}\rs^2.
\ee
The shift $\ln G$ is from the path-integral measure \cite{Bergshoeff:2019pij}. 
We denoted the beta-functionals associated with the functional couplings $E_{\mu\nu}$\,, $\CA_{\mu\nu}$\,, $\tau_\mu{}^A$, $U$\,, and $F$ as $\beta^E_{\mu\nu}$\,, $\beta^\CA_{\mu\nu}$\,, $\beta^\tau_{\mu A}$\,, $\beta^U$\,, and $\beta^F$, respectively. 
Taking \eqref{eq:id1} and \eqref{eq:id2} into account, \eqref{eq:olefa} becomes
\begin{align} \label{eq:oleare}
\begin{split}
	\Gamma_\text{1-loop} = - \frac{\log\Lambda}{4\pi\alpha'} \int_\Sigma & d^2 \sigma \, \sqrt{h} \, \Bigl[ \CD X^{A'} \, \bar{\CD} X^{B'} \bigl( \beta^E_{A'B'} + \beta^\CA_{A'B'} \bigr) + \tfrac{1}{4} \, \CD X \, \bar{\CD} \overline{X} \, \epsilon^{AB} \, \beta^\tau_{AB} \\[2pt]
	& + \tfrac{1}{2} \, \CD X \, \bar{\CD} X^{A'} \bigl( \beta^\CA_{0A'} + \beta^\CA_{1A'} \bigr) - \tfrac{1}{2} \, \CD X^{A'} \, \bar{\CD} \overline{X} \, \bigl( \beta^\CA_{0A'} - \beta^\CA_{1A'} \bigr) \\[2pt]
	& - \tfrac{1}{2} \, \lambda \, \bar{\CD} \overline{X} \, \bigl( \beta^\tau_{00} - \beta^\tau_{01} - \beta^\tau_{10} + \beta^\tau_{11} \bigr) - \lambda \, \bar{\CD} X^{A'} \, \bigl( \beta^\tau_{A'0} - \beta^\tau_{A'1} \bigr) \\[2pt]
	& - \tfrac{1}{2} \, \bar{\lambda} \, \CD X \, \bigl( \beta^\tau_{00} + \beta^\tau_{01} + \beta^\tau_{10} + \beta^\tau_{11} \bigr) - \bar{\lambda} \, \CD X^{A'} \, \bigl( \beta^\tau_{A'0} + \beta^\tau_{A'1} \bigr) \\[2pt]
	& + \lambda \, \bar{\CD} X \, \bigl( \eta^{AB} \, \beta^\tau_{AB} \bigr) + \lambda \, \bar{\lambda} \, \beta^U + \alpha' \, \text{R}^{(2)} \, \beta^F \Bigr]\,.
\end{split}
\end{align}
We introduced the notation $\CT_{A} \equiv \tau^\mu{}_{\!A} \, \CT_{\mu}$\,, $\CT_{A'} \equiv E^\mu{}_{\!A'} \, \CT_{\mu}$\,, $\CT_{(\mu\nu)} \equiv \frac{1}{2} \bigl( \CT_{\mu\nu} + \CT_{\nu\mu} \bigr)$\,, and $\CT_{[\mu\nu]} \equiv \frac{1}{2} \bigl( \CT_{\mu\nu} - \CT_{\nu\mu} \bigr)$\,. We defined $\CD X^{A'} \equiv \CD X^\mu E_\mu{}^A$, $\CD X \equiv \CD X^\mu \, \tau_\mu$\,, $\CD \overline{X} \equiv \CD X^\mu \, \bar{\tau}_\mu$\,, and analogously for expressions that involve $\bar{\CD}$\,.
The independent beta-functionals are
\begin{align} \label{eq:bfindep}
	\beta^\tau_{(AB)}
		\qquad
	\beta^\tau_{A'A}
		\qquad%
	\beta^E_{A'B'}
		\qquad%
	\beta^\CA_{AB} 
		\qquad%
	\beta^\CA_{AA'} 
		\qquad%
	\beta^\CA_{A'B'}
		\qquad%
	\beta^U
		\qquad%
	\beta^F
\end{align}
This is a direct generalization of \cite{Gomis:2019zyu, Yan:2019xsf}, now including $\beta^U$.

Furthermore, 
the local divergence $\delta^{(2)}(0)$ in \eqref{eq:ll} can be absorbed into the local dilaton counterterm. 
Also note that the beta functional $\eta^{AB} \, \beta^\tau_{AB}$ is associated with the operator $\lambda \bar{\CD} X^\mu \, \tau_\mu$ in the effective action \eqref{eq:oleare}. We focus on part of \eqref{eq:oleare},
\begin{align}
	\Gamma_\text{1-loop} & \supset - \frac{\log \Lambda}{4\pi\alpha'} \int_\Sigma d^2 \sigma \, \sqrt{h} \, \Big[ \lambda \, \bar{\CD} X^\mu \, \tau_\mu \, \bigl( \eta^{AB} \, \beta^\tau_{AB} \bigr) + \lambda \bar{\lambda} \, \beta^U \Bigr]\,.
\end{align}
Applying \eqref{eq:ll} and using the dilaton counterterm to absorb the divergence, we find
\be \label{eq:Utaucomb}
	\Gamma_\text{1-loop} \supset - \frac{\log \Lambda}{4\pi\alpha'} \int_\Sigma d^2 \sigma \, \sqrt{h} \, \lambda \bar{\lambda} \, \bigl( \beta^U - U \, \eta^{AB} \, \beta^\tau_{AB} \bigr) + \cdots\,.
\ee
Accordingly, we replace $\beta^\tau_{(AB)}$ in \eqref{eq:bfindep} with the traceless quantity,
\be \label{eq:bfindep3}
	\beta^\tau_{\{AB\}} \equiv \beta^\tau_{(AB)} - \frac{1}{2} \, \eta^{}_{AB} \, \eta^{CD} \, \beta^\tau_{CD}\,,
\ee
and identify the list of independent beta-functionals to be
\be \label{eq:bfindepf}
	\beta^\tau_{\{AB\}}
		\quad\,\,
	\beta^\tau_{A'A}
		\quad\,\,%
	\beta^E_{A'B'}
		\quad\,\,%
	\beta^\CA_{AB} 
		\quad\,\,%
	\beta^\CA_{AA'} 
		\quad\,\,%
	\beta^\CA_{A'B'}
		\quad\,\,%
	\beta^U - U \, \eta^{AB} \beta^\tau_{AB}
		\quad\,\,
	\beta^F
\ee 
This set of beta-functionals have a one-to-one correspondence to the components of the tensor coupling in front of the closed string vertex operator \eqref{eq:V1}. The fact that only the transverse components $\beta^E_{A'B'}$ of $\beta^E_{\mu\nu}$ show up in \eqref{eq:bfindepf} indicates that the sigma model \eqref{eq:SEA} is renormalizable. These beta-functionals are defined for perturbations around the Galilean-type ground state with 
\be
	E_\mu{}^{A'} = \delta_\mu^{A'}\,, 
		\qquad
	\tau_\mu{}^A = \delta_\mu^A\,, 
		\qquad
	m_\mu{}^A = U = B_{\mu\nu} = \Phi = 0\,.
\ee 

It is also manifest in \eqref{eq:SEA} that $U \rightarrow 0$ is a non-singular limit to take. This observation continues to hold at quantum level, which can be argued as follows: The Lagrangian term $\lambda \, \bar{\lambda} \, U[X]$ contains interacting operators of the form $\lambda \, \bar{\lambda} \, X^{\mu_1} \cdots X^{\mu_n} \, U_{\mu_1 \cdots \mu_n}$\,, with a coupling constant $U_{\mu_1 \cdots \mu_n}$\,. All quantum corrections from such interactions will be polynomials in the coupling constants $U_{\mu_1 \cdots \mu_n}$\,. Therefore, fine tuning the coupling constants $U_{\mu_1 \cdots \mu_n}$ to zero does not lead to any singular behavior in the resulting beta-functionals.

The beta-functionals of \eqref{eq:bfindepf} at $U = 0$ can be straightforwardly derived by applying \eqref{eq:GBP} to the standard string sigma model \eqref{eq:Srel}. 
The beta-functionals for the background fields $\hat{G}_{\mu\nu}$\,, $\hat{B}_{\mu\nu}$\,, and $\hat{\Phi}$ in \eqref{eq:Srel} are well-know \cite{Callan:1985ia}. Using the convention in \cite{Bergshoeff:2019pij}, we write these beta-functionals as
\begin{subequations} \label{eq:relbetafunctions}
\begin{align}
	\CG_{\mu\nu} \equiv {\beta}^{\hat{G}}_{\mu\nu} & = \alpha' \lr \hat{R}_{\mu\nu} + 2 \, \hat{\nabla}_{\!\mu} \hat{\nabla}_{\!\nu} \hat{\Phi} - \tfrac{1}{4} \, \hat{\mathcal{H}}_{\mu\rho\sigma} \hat{\mathcal{H}}_\nu{}^{\rho\sigma} \rr + O({\alpha'}^2)\,, \label{eq:relbetafunctionsa} \\[5pt]
	\CB_{\mu\nu} \equiv {\beta}^{\hat{B}}_{\mu\nu} & = \alpha' \lr - \tfrac{1}{2} \hat{\nabla}^\rho \hat{\mathcal{H}}_{\rho\mu\nu} + \hat{\nabla}^\rho \hat{\Phi} \, \hat{\mathcal{H}}_{\rho\mu\nu} \rr + O({\alpha'}^2)\,, \label{eq:relbetafunctionsb} \\[5pt]
	\CF \equiv {\beta}^{\hat{F}} & =
		- \alpha' \lr \hat{\nabla}_{\!\mu} \hat{\nabla}^\mu \hat{\Phi} + \tfrac{1}{4} \hat{R} - \hat{\nabla}_{\!\mu} \hat{\Phi} \, \hat{\nabla}^\mu \hat{\Phi} - \tfrac{1}{48} \, \hat{\mathcal{H}}_{\mu\nu\lambda} \hat{\mathcal{H}}^{\mu\nu\lambda} \rr + O({\alpha'}^2)\,,
\end{align}
\end{subequations}
where $\hat{\CH}_{\mu\nu\rho} \equiv \p_\mu \hat{B}_{\nu\rho} + \p_\nu \hat{B}_{\rho\mu} + \p_\rho \hat{B}_{\mu\nu}$ and $\hat{F} \equiv \hat{\Phi} - \frac{1}{4} \ln (- \hat{G})$\,, with $\hat{G}$ the determinant of $\hat{G}_{\mu\nu}$\,. We already assumed that we are in the critical dimensions. Since we are ultimately interested in the limit $U \rightarrow 0$\,, we will omit any derivatives acting on $U$ in the calculation. This is based on the following observation: although $U$ appears in denominators (and in $\ln U$) in \eqref{eq:GBP} and should be treated carefully in the $U \rightarrow 0$ limit, derivatives of $U$ only appear in numerators and can be immediately set to zero.
In the regime where $U$ is sufficiently small, the following expansions with respect to $U$ have been worked out in \cite{Bergshoeff:2021bmc}:
\begin{subequations} \label{eq:relations}
\begin{align} 
	\CG_{A}{}^A - \epsilon^{AB} \, \CB_{AB} & = \alpha' \, \langle S_+ \rangle + O (U)\,, 
		&%
	\CG_{\{AB\}} & = \alpha' \, U^{-1} \, \langle G \rangle_{\{AB\}} + O(U)\,, \\[2pt]
	\CG_{A}{}^A + \epsilon^{AB} \, \CB_{AB} & = \alpha' \, U^{-2} \, \langle S_- \rangle + O (U^{-1})\,,
		&%
	\CG_{A'B'} & = \alpha' \, \langle G \rangle_{A'B'} + O(U)\,, \\[2pt]
	\CG_{AA'} - \epsilon_A{}^B \CB_{BA'} & = \alpha' \, U^{-1} \, \langle V_+ \rangle_{AA'} + O(1)\,, 
		&%
	\CB_{A'B'} & = \alpha' \, \langle B \rangle_{A'B'} + O(U)\,, \\[2pt]
	\CG_{AA'} + \epsilon_A{}^B \CB_{BA'} & = \alpha' \, \langle V_- \rangle_{AA'} + O(U)\,, 
		&%
	\CF & = - \alpha' \, \langle \Phi \rangle + O(U)\,.
\end{align}
\end{subequations}
Note that the $A$ and $A'$ indices in $\CG$ and $\CB$ come from contracting the curved index $\mu$ with $\tau^\mu{}_{\!A}$ and $E^\mu{}_{\!A'}$\,, respectively.~\footnote{\,Our convention is slightly different from \cite{Bergshoeff:2021bmc}, where the curved index $\mu$ is contracted with the relativistic inverse Vielbein field $\hat{E}^\mu{}_M$\,.} The notation $\CG_{\{AB\}}$ denotes the traceless part of $\CG_{AB}$\,, analogous to the definition in \eqref{eq:bfindep3}. Here, $\langle S_\pm \rangle$\,, $\langle V_\pm \rangle_{AA'}$\,, $\langle G \rangle_{\{AB\}}$\,, $\langle G \rangle_{A'B'}$\,, $\langle B \rangle_{A'B'}$\,, and $\langle \Phi \rangle$ are given in Eqn.~(49) and (54) of \cite{Bergshoeff:2021bmc}. The detailed expressions of these symbols are not important here, and we simply need to note that they are independent of $U$\,. We emphasize that \eqref{eq:relations} holds only under the condition $\p_\mu U = 0$\,.  
Moreover, using \eqref{eq:GBP} we find
\begin{subequations} \label{eq:GBF}
\begin{align}
	\CG_{\mu\nu} & = - U^{-2} \, \tau^{}_{\mu\nu} \, \beta^U + U^{-1} \, \bigl( \tau^{}_\mu{}^A \, \beta^\tau_{\nu A} +\tau^{}_\nu{}^A \, \beta^\tau_{\mu A} \bigr) + \beta^E_{\mu\nu}\,, \\[2pt]
	\CB_{\mu\nu} & = U^{-2} \, \epsilon_{AB} \, \tau^{}_\mu{}^A \, \tau^{}_\nu{}^B \, \beta^U - U^{-1} \, \bigl( \tau^{}_\mu{}^A \, \beta^\tau_{\nu B} - \tau^{}_\nu{}^A \, \beta^\tau_{\mu B} \bigr) \, \epsilon_A{}^B + \beta^\CA_{\mu\nu}\,, \\[3pt]
	\CF & = \beta^\Phi - \tfrac{1}{2} \, U^{-1} \, \beta^U - \tfrac{1}{4} \, \beta^E_{A'A'} - \tfrac{1}{4} \, U \, \tau^{\mu\nu} \, \CG_{A}{}^A
		= \beta^F\,.
\end{align}
\end{subequations}
We already defined $F$ in \eqref{eq:defF}, which implies that
$\beta^F = \beta^\Phi - \tfrac{1}{4} \, \beta^E_{A'A'} - \tfrac{1}{2} \, \eta^{AB} \, \beta^\tau_{AB}$\,.
From \eqref{eq:GBF}, we find
\begin{subequations} \label{eq:GBFinv}
\begin{align}
	\beta^\tau_{\{AB\}} & = \tfrac{1}{2} \, U \, \CG_{\{AB\}}\,, 
		&%
	\beta^\CA_{AB} & = \tfrac{1}{2} \, \epsilon_{AB} \, \bigl( \CG_C{}^C - \epsilon^{CD} \, \CB_{CD} \bigr)\,, 
		&
	\beta^F & =  \CF\,, \\[2pt]
	\beta^\tau_{A'A} & = U \, \CG_{AA'}\,, 
		&%
	\beta^\CA_{AA'} & = \epsilon_A{}^B \, \bigl( \CG_{BA'} + \epsilon_B{}^C \, \CB_{CA'} \bigr)\,, \\[2pt]
	\beta (E)_{A'B'} & = \CG_{A'B'}\,, 
		&%
	\beta^\CA_{A'B'} & = \CB_{A'B'}\,, 
\end{align}
\end{subequations}
and
\begin{align} \label{eq:comUtau}
	\beta^U - U \, \eta^{AB} \, \beta^\tau_{AB} = - \tfrac{1}{2} \, U^2 \, \CG_A{}^A. 
\end{align}
Note that $\beta^U$ and $\eta^{AB} \, \beta^\tau_{AB}$ appear in the same combination that we found earlier in \eqref{eq:Utaucomb}. The linearized beta-functionals in perturbations around flat background fields are consistent with the results in \cite{Gomis:2019zyu}.

As we have argued earlier, the beta-functionals in \eqref{eq:bfindepf} are finite in the limit $U \rightarrow 0$\,. Plugging \eqref{eq:relations} into \eqref{eq:GBFinv} and \eqref{eq:comUtau}, we find, at the lowest order in $\alpha'$,
\begin{subequations} \label{eq:U0bf}
\begin{align}
	\beta^\tau_{\{AB\}} \Big|_{U=0} \!\! & = \tfrac{1}{2} \, \alpha' \, \langle G \rangle_{\{AB\}}\,, 
		&\!%
	\beta^\CA_{AB} \Big|_{U=0} \!\! & = \tfrac{1}{2} \, \alpha' \, \epsilon_{AB} \, \langle S_+ \rangle\,, 
		&\!%
	\beta^F \Big|_{U=0} \!\! & = - \alpha' \,  \langle \Phi \rangle\,, \\[2pt]
	\beta^\tau_{A'A} \Big|_{U=0} \!\! & = \tfrac{1}{2} \, \alpha' \, \langle V_+ \rangle_{AA'}\,, 
		&\!%
	\beta^\CA_{AA'} \Big|_{U=0} \!\! & = \alpha' \, \epsilon_A{}^B \, \langle V_- \rangle_{BA'}\,, \\[2pt]
	\beta (E)_{A'B'} \Big|_{U=0} \!\! & = \alpha' \, \langle G \rangle_{A'B'}\,, 
		&\!%
	\beta^\CA_{A'B'} \Big|_{U=0} \!\! & = \alpha' \, \langle B \rangle_{A'B'}\,, 
\end{align}
\end{subequations}
and
\be \label{eq:betaU}
	\beta^U \Big|_{U=0} = - \tfrac{1}{4} \, \alpha' \, \langle S_- \rangle = \alpha' \, T^{}_{A'B'} \, \overline{T}^{}_{\!A'B'} \,.
\ee
We have defined $T_{\mu\nu}{}^A \equiv \p_{[\mu} \tau_{\nu]}{}^A$, and accordingly $T_{\mu\nu} \equiv \p_{[\mu} \tau_{\nu]}$ and $\overline{T}_{\mu\nu} \equiv \p_{[\mu} \bar{\tau}_{\nu]}$\,.
Note that $\beta^\tau_{AB}$\,, which is finite at $U = 0$\,, drops out in \eqref{eq:betaU}. We have spelled out the explicit expression for $\langle S_- \rangle$, which can be found in Eqn.~(49f) of \cite{Bergshoeff:2021bmc}. 
The expression of $\beta^U$ in \eqref{eq:betaU} corroborates the results in \cite{Gomis:2019zyu, Yan:2019xsf, Gallegos:2019icg}, where the quantum calculation is performed using \eqref{eq:SEA}. Imposing Weyl invariance sets the beta-functionals to zero and gives rise to the target-space equations of motion in \cite{Bergshoeff:2021bmc, Gallegos:2020egk} that determine the spacetime dynamics at $U=0$\,, describing a nonrelativistic corner of string theory. In the following discussions, we will refer to the beta-functionals in \eqref{eq:U0bf} and \eqref{eq:betaU} with the condition $U = 0$ imposed implicitly.

Following similar arguments as in \cite{Gomis:2019zyu, Yan:2019xsf}, we give a nonrenormalization argument in Appendix \ref{app:nonre} that shows $\beta^U$ vanishes at all loops if~\footnote{\,In a footnote of \cite{Gomis:2020fui}, there is a preliminary comment suggesting that the condition $T_{A'B'} \overline{T}_{A'B'} = 0$ from the vanishing beta-functional $\beta^U = 0$ at the lowest order in $\alpha'$ might be sufficient for $\beta^U = 0$ at higher orders in $\alpha'$. This statement does not seem to hold after a more careful examination. Instead, a stronger condition \eqref{eq:dt0} is required for the nonrenormalization argument in Appendix \ref{app:nonre}.} 
\be \label{eq:dt0}
	T_{\mu\nu} = - \Omega_{[\mu} \tau_{\nu]}
		\quad \text{or} \quad
	\overline{T}_{\mu\nu} = \Omega_{[\mu} \bar{\tau}_{\nu]}
\ee 
holds. Here, $\Omega_\mu$ denotes the longitudinal spin connection. 
The condition \eqref{eq:dt0} implies that $T_{A'B'} = 0$ or $\overline{T}_{A'B'} = 0$\,, which indeed sets $\beta^U$ to zero at the lowest order in $\alpha'$ in \eqref{eq:betaU}. In addition, the condition \eqref{eq:dt0} also gives rise to an extra geometric constraint $\tau^\mu\, \overline{T}_{\mu A'} = 0$\,, with $\tau^\mu \equiv \frac{1}{2} \left( \tau^\mu{}_0 + \tau^\mu{}_1 \right)$\,. This extra constraint is needed for the nonrenormalization argument in Appendix \ref{app:nonre}. In this regard, the torsional constraints imposed by setting $\beta^U = 0$ at the lowest $\alpha'$ in \eqref{eq:betaU} might not be sufficient for higher-loop contributions in $\beta^U$ to vanish. According to \eqref{eq:betaU}, there may exist solutions to the vanishing beta-functionals at the lowest $\alpha'$ that violate the geometric constraint $\tau^\mu \, \overline{T}_{\mu A'} = 0$\,. If there are higher-loop contributions to $\lambda\bar{\lambda}$ that also require $\tau^\mu \, \overline{T}_{\mu A'} = 0$ for the associated beta-functional to vanish, then, for the solutions that violate such a geometric constraint to survive, a nonzero $U$ is needed at higher loops.~\footnote{\,Note that the geometric constraint $\tau^\mu \, \overline{T}_{\mu A'} = 0$ cannot be recovered in a perturbative manner with respect to $\alpha'$.} This would deform the theory towards relativistic string theory.

The above observation that the $\lambda\bar{\lambda}$ operator is not renormalized under the condition in \eqref{eq:dt0} strongly suggest that there exists a notion of self-contained nonrelativistic string theory defined by a renormalizable worldsheet QFT, and begs for a symmetry reasoning that underlies the condition \eqref{eq:dt0}. One related symmetry argument has been given in \cite{Gomis:2019zyu, Yan:2019xsf}, which we review now. Note that the geometric constraints in \eqref{eq:dt0} can be embedded within the zero-torsion condition
$T_{\mu\nu}{}^A = \epsilon^{A}{}_B \, \Omega_{[\mu} \tau_{\nu]}{}^B$\,. These torsional constraints preserve the two-dimensional foliation structure in the spacetime geometry. 
When the zero-torsion condition is imposed, the (divergent and finite) quantum corrections to $\lambda \bar{\lambda}$ vanish identically at all loop orders and $U=0$ is protected \cite{Gomis:2019zyu, Yan:2019xsf}. It is in this sense that we refer to the $\lambda\bar{\lambda}$ term as the torsional deformation, as we have preluded in the previous section. From the spacetime point of view, certain torsional constraints are needed in order to define a genuinely nonrelativistic geometry equipped with a robust foliation structure. 

When the zero-torsion condition is satisfied, there is an emergent $Z_A$ gauge symmetry that protects $U=0$ \cite{Gomis:2019zyu, Yan:2019xsf}, which we will detail in \S\ref{sec:netc}. The associated symmetry algebra is referred to as the string Newton-Cartan algebra, applying which to the sigma model determines the appropriate spacetime geometry that nonrelativistic string theory is coupled to. This spacetime geometry is string Newton-Cartan geometry with the zero-torsion condition.  However, it is suggested in \cite{Bergshoeff:2021bmc} that zero-torsion constraint might be too strong to be generalized to the supersymmetric case. Moreover, since the weaker conditions in \eqref{eq:dt0} already seem to protect $U$ from receiving any quantum correction, one may suspect that there is a modified notion of string Newton-Cartan symmetries that breaks part of the $Z_A$ symmetry, which we explore next. 

\subsection{Noncentral Extensions and Torsional Constraints} \label{sec:netc}

We now return to the examination of nonrelativistic symmetry algebras that underly different string sigma models. The symmetries of the free theory have been classified in \eqref{eq:globaltrnsf}. Generically, not all the infinite-dimensional symmetries of the free theory in \eqref{eq:globaltrnsf} are preserved when interactions are turned on. In the following, we discuss three subalgebras of \eqref{eq:globaltrnsf} that are realized as the symmetry algebras underlying different sigma models that describe strings propagating in non-Lorentzian geometries. 

\subsubsection*{String Galilei Symmetries} 

First, fine-tuning the physical value of the marginal coupling $U$ to zero in \eqref{eq:SEA} leads to the classical action
\begin{align} \label{eq:zeroUaction}
\begin{split}
	S \! = \! \frac{1}{4\pi\alpha'} \! \int_\Sigma \! d^2 \sigma \sqrt{h} \Bigl[ \CD X^\mu \, \bar{\CD} X^\nu \bigl( E_{\mu\nu} \! + \! \CA_{\mu\nu} \bigr)
		+ \lambda \, \bar{\CD} X^\mu \tau_\mu + \bar{\lambda} \, \CD X^\mu \bar{\tau}_\mu + \alpha' \, \text{R}^{(2)} \, \Phi \Bigr],
\end{split}
\end{align}
where $\CA_{\mu\nu} = B_{\mu\nu} + \bigl( m_\mu{}^A \, \tau_\nu{}^B - m_\nu{}^A \, \tau_\mu{}^B \bigr) \, \epsilon_{AB}$\,, with $B_{\mu\nu}$ being the Kalb-Ramond field. 
Classically, the sigma model \eqref{eq:zeroUaction} has a nonrelativistic string spectrum and the target space geometry is described by the string Newton-Cartan data $E_\mu{}^{A'}$, $\tau_\mu{}^A$, and $m_\mu{}^A$, whose gauge transformations are \cite{Bergshoeff:2021bmc},
\begin{subequations} \label{eq:trnsfG}
\begin{align}
	\delta \tau_\mu{}^A & = \Lambda_\text{D} \, \tau_\mu{}^A + \Lambda \, \epsilon^A{}_B \, \tau_\mu{}^B\,, \label{eq:trnsfGtau} \\[2pt]
	\delta E_\mu{}^{A'} & = \Lambda^{A'}{}_{\!A} \, \tau_\mu{}^A + \Lambda^{A'}{}_{\!B'} \, E_\mu{}^{B'}, \\[2pt]
	\delta m_\mu{}^A & = - \Lambda_\text{D} \, m_\mu{}^A + \Lambda \, \epsilon^A{}_B \, m_\mu{}^A + \Lambda^A{}_{\!A'} \, E_\mu{}^{A'},
\end{align}
\end{subequations}
which can be read off from \eqref{eq:deltasnc} and \eqref{eq:mmuA} by setting $U = 0$\,.
Here, the full Lorentz boost is absent, and the boost transformation $\Lambda^{A'}{}_A$ only acts on the transverse Vielbein field $E_\mu{}^{A'}$ but not the longitudinal Vielbein field $\tau_\mu{}^A$. This broken boost transformation is referred to as the string Galilei boost. Moreover, the dilaton transforms as a scalar with a dilatational charge; infinitesimally, we have $\delta \Phi = \Lambda_\text{D} \Phi$\,.  
The infinitesimal transformations in \eqref{eq:trnsfG} form the string Galilei algebra defined in Appendix \ref{app:sga}{\color{blue}.2}. 

Quantum mechanically, the $\lambda\bar{\lambda}$ term is generated because of the nontrivial beta-functional \eqref{eq:betaU}. This implies that \eqref{eq:zeroUaction} is not renormalizable and a $\lambda \bar{\lambda}$ counterterm has to be added in order to cancel the divergent quantum corrections, such that the beta-functional \eqref{eq:betaU} can be defined after imposing appropriate renormalization conditions. Therefore, the string Galilei symmetries generated by \eqref{eq:trnsfG} are not sufficient for protecting \eqref{eq:zeroUaction} from the torsional deformation towards relativistic string theory.

\subsubsection*{String Newton-Cartan Symmetries} 

In \cite{Andringa:2012uz, Bergshoeff:2019pij}, noncentral extensions of the string Galilei algebra are studied, which leads to a larger subalgebra of the infinite-dimensional algebra \eqref{eq:globaltrnsf}. This is dubbed as the string Newton-Cartan algebra, which has been realized as the symmetry algebra underlying \eqref{eq:zeroUaction}, together with the zero-torsion constraint in the target space.  

In string Galilei algebra, the string Galilei boost generator $G_{AA'}$ and the transverse translational generator $P_{A'}$ commute.\,\footnote{\,In the Poincar\'{e} algebra, the boost and transverse translational generator would commute into the longitudinal translational generator.} As a result, in \eqref{eq:trnsfGtau}, $\tau_\mu{}^A$ does not transform into $E_\mu{}^{A'}$ under the string Galilei boost. The string Galilei algebra can be extended by requiring that $G_{AA'}$ and $P_{A'}$ commute into a new generator $Z_A$\,, with
\be \label{eq:noncen}
	[G_{AA'}\,, P_{B'}] = \delta_{A'B'} \, Z_A\,.
\ee
Gauging $Z_A$ will lead to the gauge field $m_\mu{}^A$ that we introduced as a convenient parametrization of the background fields thus far \cite{Andringa:2012uz}. Note that $Z_A$ is a noncentral extension; for example, the commutation relation between $Z_A$ and the longitudinal Lorentz boost generator $M$ is nontrivial. 
The $Z_A$ transformation, which we parametrize by $\sigma^A$\,, only acts nontrivially on $m_\mu{}^A$\,, with 
\be \label{eq:deltam}
	\delta^{}_{Z} m_\mu{}^A = \p_\mu \sigma^A - \epsilon^A{}_B \, \sigma^B \, \Omega_\mu \,.
\ee
Here, $\Omega_\mu$ denotes the longitudinal spin connection.
Requiring that \eqref{eq:deltam} be a symmetry transformation in \eqref{eq:zeroUaction} automatically leads to the zero-torsion condition \cite{Andringa:2012uz, Bergshoeff:2019pij},
\be \label{eq:ztcs}
	T_{\mu\nu}{}^A = \epsilon^A{}_B \, \Omega_{[\mu} \tau_{\nu]}{}^B\,.
\ee 
This zero-torsion condition imposes the geometric constraints
\be \label{eq:gc}
	T^{}_{A'(AB)} = T^{}_{A'B'A} = 0\,.
\ee
The remaining constraints from \eqref{eq:ztcs} can be used to solve for the spin connection $\Omega_\mu$ \cite{Andringa:2012uz, Bergshoeff:2019pij}.
The full symmetry algebra that contains this $Z_A$ symmetry is the string Newton-Cartan algebra, which we present in Appendix \ref{app:sNCa}{\color{blue}.3}. 
Note that the longitudinal dilatational symmetry parametrized by $\Lambda_\text{D}$ in \eqref{eq:trnsfG} is not preserved by the constraints in \eqref{eq:gc}, unless the extra condition $E^\mu{}_{\!A'} \, \p_\mu \Lambda_\text{D} = 0$ is imposed \cite{Bergshoeff:2019pij}. 

Since the $\lambda \bar{\lambda}$ operator simply does not exist, we can no longer combine the beta-functional $\eta^{AB} \beta^\tau_{AB}$ with $\beta^U$ as in \eqref{eq:Utaucomb}. In contrast, $\eta^{AB} \beta^\tau_{AB}$ becomes independent \cite{Gomis:2019zyu}. Nevertheless, nonrenormalization theorems proven in \cite{Yan:2019xsf} show that all the beta-functionals associated with $\tau_\mu{}^A$ are trivially zero at all loops due to the $Z_A$ symmetry. In practice, the nontrivial beta-functionals at $U=0$ are $\beta^E_{A'B'}$\,, $\beta^\CA_{\mu\nu}$\,, and $\beta^F$ in \eqref{eq:U0bf}. Because we do not have a counterterm associated with the $\lambda\bar{\lambda}$ operator, in this auxiliary limiting procedure, $U$ should be treated as an external parameter that does not receive any RG flow. Moreover, the constraints in \eqref{eq:gc} have to be imposed before the $U \rightarrow 0$ limit is even considered.~\footnote{\,This can be understood by the following argument: We want to compute the beta-functionals using the sigma model \eqref{eq:zeroUaction} that realize the string Newton-Cartan symmetries, which means that the constraints in \eqref{eq:gc} are already imposed. As a technical trick, we first turn on a small constant $U = U_0$ as an auxiliary parameter (but without any extra counterterm) and then evaluate the quantum corrections. This is equivalent to introducing a Gaussian smearing of the constraints imposed by integrating out $\lambda$ and $\bar{\lambda}$ in the path integral at $U_0 = 0$\,. At the end of the calculation, we set $U_0 = 0$\,. This procedure does not change the effective action. Since we now have an auxiliary $\lambda\bar{\lambda}$ operator in the sigma model, the quantum calculation can be done equivalently by first integrating out $\lambda$ and $\bar{\lambda}$ in the path integral. This of course leads to the standard relativistic sigma model, whose beta-functionals are given in \eqref{eq:relbetafunctions}. However, since we assumed that there is no $\lambda\bar{\lambda}$ counterterm and imposed the constraints in \eqref{eq:gc}, the same conditions have to be applied to \eqref{eq:relbetafunctions} as well. This means that $\beta^U$ has to be eliminated in \eqref{eq:relbetafunctions}. Finally, the desired beta-functionals for the nonrelativistic string sigma model can be uncovered by setting $U_0 = 0$\,, which is required by self-consistency. \label{ft:bf}} The appropriate limit of the relativistic beta-functionals that lead to the ones of nonrelativistic string sigma model with string Newton-Cartan symmetries has been worked out in \cite{Bergshoeff:2019pij}. This reproduces the results found by explicit quantum calculations using the action \eqref{eq:zeroUaction} \cite{Gomis:2019zyu, Yan:2019xsf}. 
With the string Newton-Cartan symmetries imposed, we have a well-defined notion of nonrelativistic string theory that forbids any deformation towards relativistic string theory, at least when spacetime EFTs are concerned. The spacetime dynamics of the EFTs is determined by the renormalizable worldsheet QFT.

\subsubsection*{Halve the Noncentral Extension} 

We already noted that the zero-torsion condition \eqref{eq:ztcs} is not invariant under the longitudinal dilatational symmetry. 
Is there a way to develop a symmetry argument that prohibits the $\lambda \bar{\lambda}$ operator from being generated quantum mechanically in \eqref{eq:zeroUaction}, while preserving the longitudinal dilatational symmetry? This can be achieved by breaking half of the $Z_A$ symmetry. In the meantime, requiring that the longitudinal Lorentz boosts be preserved motivates us to break a lightlike part of the $Z_A$ symmetry. Define $Z \equiv Z_0 + Z_1$ and $\overline{Z} \equiv Z_0 - Z_1$\,. We choose to break the $\overline{Z}$ symmetry by taking the contraction $\overline{Z} \rightarrow 0$ in the string Newton-Cartan algebra.~\footnote{\,One may instead choose to keep the $\overline{Z}$ symmetry and break the $Z$ symmetry, which does not make any physical difference.} This defines a self-consistent subalgebra that we present in Appendix \ref{app:msNCa}{\color{blue}.4}. As a result, the commutation relation in \eqref{eq:noncen} becomes
\be 
	[G_{A'}\,, P_{B'}] = \delta_{A'B'} \, Z\,,
		\qquad
	[\bar{G}_{A'}\,, P_{B'}] = 0\,,
\ee
where $G_{A'} \equiv G_{0A'} + G_{1A'}$ and $\bar{G}_{A'} \equiv G_{0A'} - G_{1A'}$\,. Parametrizing the $Z$ transformation by $\sigma$\,,~\footnote{\,This Lie group parameter $\sigma$ should not be confused with the worldsheet coordinate $\sigma$\,.} we find that the only nontrivial transformation under the action of $Z$ is
\be \label{eq:deltaZm}
	\delta^{}_Z \, m_\mu = \bigl( \p_\mu + \Omega_\mu \bigr) \sigma\,.
\ee
We have defined $m_\mu \equiv m_\mu{}^0 + m_\mu{}^1$ and $\bar{m}_\mu \equiv m_\mu{}^0 - m_\mu{}^1$\,, with $\delta^{}_Z \, \bar{m}_\mu = 0$\,. Requiring that \eqref{eq:deltaZm} be invariant under \eqref{eq:zeroUaction}, we find the following torsional constraint:
\be \label{eq:taubarc}
	\overline{T}_{\mu\nu} = \Omega_{[\mu} \bar{\tau}_{\nu]}\,,
\ee
which coincides with the condition \eqref{eq:dt0}, under which the $\lambda\bar{\lambda}$ operator is protected against quantum corrections. This statement is also evident from the nonrenormalization arguments in Appendix \ref{app:nonre}. The condition \eqref{eq:taubarc} leads to the geometric constraints,
\begin{align} \label{eq:gcsusy}
	\tau^\mu \, \overline{T}_{\mu A'} = 0\,,
		\qquad
	\overline{T}_{A'B'} = 0\,.
\end{align}
or, more explicitly, $E^\mu{}_{\!A'} \, \tau^\nu \, \p_{[\mu} \bar{\tau}_{\nu]} = E^\mu{}_{\!A'} \, E^\nu{}_{\!B'} \, \p_{[\mu} \bar{\tau}_{\nu]} = 0$\,. Here, $\tau^\mu \equiv \frac{1}{2} \bigl( \tau^\mu{}_0 + \tau^\mu{}_1 \bigr)$\,.
According to \eqref{eq:betaU}, the second condition in \eqref{eq:gcsusy} already suffices for the $\lambda\bar{\lambda}$ operator to be free from quantum corrections at the lowest order in $\alpha'$. In addition, we now also have an extra geometric constraint $\tau^\mu \, \overline{T}^{}_{\mu A'} = 0$\,. In Appendix \ref{app:nonre}, we use Feynman diagrams to show that the constraints in \eqref{eq:gcsusy} are sufficient for higher-loop quantum corrections of $\lambda\bar{\lambda}$ to vanish, which is expected from our symmetry argument.
These constraints in \eqref{eq:gcsusy} restrict the foliation structure and still allow nonzero torsions. 
Moreover, they are compatible with the longitudinal dilatational symmetry parametrized by $\Lambda_\text{D}$\,. Fascinatingly, \eqref{eq:gcsusy} are precisely the constraints found in \cite{Bergshoeff:2021bmc}, by requiring that the supersymmetry rules are finite in a nonrelativistic limit of heterotic supergravity . 

Since the $\lambda \bar{\lambda}$ operator is absent, the beta-functional $\eta^{AB} \, \beta^\tau_{AB}$ cannot be combined with $\beta^U$ as in \eqref{eq:Utaucomb} and now becomes independent. 
To acquire the correct set of independent beta-functionals, we first impose the constraints \eqref{eq:gcsusy} in \eqref{eq:relbetafunctions} (under the parametrizations in \eqref{eq:GBP}), and then take the nonsingular limit $U \rightarrow 0$\,.~\footnote{\,See Footnote \ref{ft:bf} for justifications.}  We emphasize that $U = U_0$ is a constant auxiliary parameter that does not receive any RG flow.
While the relations in \eqref{eq:GBFinv} continue to hold but now with the constraints in \eqref{eq:gcsusy} being imposed, \eqref{eq:comUtau} is modified to be
\be \label{eq:betatautwistless}
	\eta^{AB} \, \beta^\tau_{AB} = \tfrac{1}{2} \, U_0 \, \CG_A{}^A\,,
\ee
where $\CG_A{}^A$ is given in \eqref{eq:relations}, with
\be
	\CG_A{}^A = \tfrac{1}{2} \, \alpha' \, U_0^{-2} \langle S_- \rangle + 2 \, \alpha' \, U_0^{-1} \, \langle Q \rangle + O(U_0^{-2})\,.
\ee
Here, $\langle S_- \rangle$ is given in \eqref{eq:betaU}. Taking \eqref{eq:gcsusy} into account, we have $\langle S_- \rangle = 0$ and
\begin{align}
\begin{split}
	\langle Q \rangle & = - \tau^\mu \bigl( \p_{A'} T_{A'\mu} 
		- \tfrac{1}{2} \, T_{A'B'} \, \CF_{A'B'\mu} + 2 \, T_{\mu A'} \, \p_{A'} \Phi \bigr) \\[2pt] 
		& \quad - \bar{\tau}^\mu \bigl( \p_{A'} \overline{T}_{A'\mu} + 2 \, \overline{T}_{\mu A'} \, \p_{A'} \Phi \bigr) + \text{covariantizations}.
\end{split}
\end{align}
We defined $\CF_{\mu\nu\rho} = \p_\mu \CA_{\nu\rho} + \p_\rho \CA_{\mu\nu} + \p_\nu \CA_{\rho\mu}$\,. 
Plugging all the ingredients back into \eqref{eq:betatautwistless}, we find
\be \label{eq:betatauresult}
	\eta^{AB} \, \beta^\tau_{AB} \Big|_{U = 0} = \alpha' \, \langle Q \rangle\,.
\ee
At the linearized order, \eqref{eq:betatauresult} reproduces the result derived by evaluating OPEs in \cite{Gomis:2019zyu}.
The remaining beta-functionals are the same as the ones in \eqref{eq:U0bf}, now with the torsional constraints \eqref{eq:gcsusy} taken into account.

In this framework where the noncentral extension is halved, we are still able to achieve a self-contained notion of nonrelativistic string theory defined by a (local) renormalizable worldsheet QFT, which describes strings propagating in a torsional string Newton-Cartan geometry. In particular, the torsional deformation $\lambda\bar{\lambda}$ is strictly prohibited. It is promising that this new symmetry group defines the appropriate spacetime geometry that is extendable to a Galilean-type supergravity.

\acknowledgments

The author would like to thank Eric~A.~Bergshoeff, Jaume Gomis, Johannes Lahnsteiner, Luca Romano, Jan Rosseel, Ceyda \c{S}im\c{s}ek, and Matthew Yu for numerous discussions that partly inspired this work. It is also a great pleasure to thank Niels A. Obers, Gerben Oling, Troels Harmark, Jelle Hartong, Umut G\"{u}rsoy, Laurent Freidel, Djordje Minic, and Leo Bidussi for stimulating discussions.

\appendix

\section{Longitudinal T-duality} \label{app:Tdual}

In this appendix, following the discussions in \S\ref{sec:ofdc}, we provide an equivalent analysis of the longitudinal T-duality transformations along both the longitudinal space and time coordinates. This will make manifest the connection to the discrete light cone quantization (DLCQ) of relativistic string theory \cite{Bergshoeff:2018yvt}.
We first take the T-dual of \eqref{eq:cg} along the $X^1$-direction by rewriting the action in the equivalent form,
\begin{align} \label{eq:Sparent}
	S_\text{parent} & \! = \! \frac{1}{4\pi\alpha'} \! \int \! d^2 \sigma \Bigl[ \p_\alpha X^{A'} \p^\alpha X^{A'} \! + \! \lambda \, \bigl( \bar{\p} X^0 \! + \bar{v} \bigr) + \! \bar{\lambda} \, \bigl( \p X^0 \! - v \bigr) + Y_1 \bigl( \bar{\p} v - \p \bar{v} \bigr)\Bigr]. 
\end{align}
Integrating out the auxiliary field $Y_1$ imposes $\bar{\p} v = \p \bar{v}$\,, which can be solved locally by $v = \p X^1$\,, $\bar{v} = \bar{\p} X^1$\,.
This gives back the original action \eqref{eq:cg}. To pass on to the T-dual frame, we instead integrate out $v$ and $\bar{v}$\,, which induces the relations $\lambda = - \p Y_1$ and $\bar{\lambda} = - \bar{\p} Y_1$\,,
with $Y_1$ gaining the interpretation of the dual coordinate compactified over a circle of radius $\alpha' / R$\,. The dual action is
\be \label{eq:dual}
	\tilde{S}_0 = \frac{1}{4\pi\alpha'} \int_\Sigma d^2 \sigma \, \Bigl( \p_\alpha X^{A'} \p^\alpha X^{A'} - \p Y_1 \, \bar{\p} X^0 - \bar{\p} Y_1 \, \p X^0 \Bigr) \,,
\ee
which describes relativistic string theory with a lightlike circle in the $Y_1$-direction. This is the DLCQ of string theory \cite{Motl:1997th, Banks:1996my, Dijkgraaf:1997vv}. 

We now perform a second T-duality transformation along the $X^0$-direction in \eqref{eq:dual}. Note that the $X^0$-circle is lightlike now. We start with the equivalent action,
\begin{align} \label{eq:Sp0}
	\tilde{S}_\text{parent} & = \frac{1}{4\pi\alpha'} \int_\Sigma d^2 \sigma \, \Bigl[ \p_\alpha X^{A'} \p^\alpha X^{A'} - \p Y_1 \, \bar{u} - \bar{\p} Y_1 \, u + Y_0 \bigl( \bar{\p} u - \p \bar{u} \bigr) \Bigr]\,.
\end{align}
Integrating out $Y^0$ in \eqref{eq:Sp0} imposes $\bar{\p} u = \p \bar{u}$\,, which is solved locally by $u = \p X^0$\,, $\bar{u} = \bar{\p} X^0$.
This sets $\tilde{S}_\text{parent} = \tilde{S}_0$\,. 
Instead, 
rewriting \eqref{eq:Sp0} as
\be \label{eq:ddaction}
	S'_0 = \frac{1}{4\pi\alpha'} \int_\Sigma d^2 \sigma \, \Bigl( \p_\alpha X^{A'} \p^\alpha X^{A'} - u \, \bar{\p} Y + \bar{u} \, \p \overline{Y} \Bigr)\,,
\ee
where $Y = Y_0 + Y_1$ and $\overline{Y} = Y_0 - Y_1$\,, we find that $u$ and $\bar{u}$ are Lagrange multipliers that impose the (anti-)holomorphic conditions $\bar{\p} Y = \p \overline{Y} = 0$\,. The action $S'_0$ in \eqref{eq:ddaction} describes nonrelativistic string theory with the spacetime coordinates $(Y_A, X^{A'})$\,, with $Y_A$ dual to $X^A$. 

\section{Hamiltonian Formalism and Generalized Metric} \label{app:Hfgm}

In the course of understanding the spacetime geometry and the spacetime EFT in nonrelativistic string theory, the formalism in \eqref{eq:zeroUaction} that involves string Newton-Cartan geometry plays an important role. On the other hand, the interplay between string Newton-Cartan geometry, the Kalb-Ramond and dilaton backgrounds reveals abundant redundancies parametrized by the Stueckelberg symmetries in \eqref{eq:stueckelberg}. It is therefore motivating to find a formalism in which all the background fields are manifestly invariant under these Stueckelberg symmetries. This is indeed possible by passing on to the Hamiltonian formalism. 

We start with the action in \eqref{eq:zeroUaction}. For simplicity, we work in flat worldsheet, which means that the following discussion will not involve the dilaton term. Setting $\alpha' = 1/2$\,, the action \eqref{eq:zeroUaction} now reads
\begin{align} \label{eq:S}
	S = - \frac{1}{4\pi\alpha'} \int_\Sigma d^2 \sigma \Bigl[ \p X^\mu \, \bar{\p} X^\nu \bigl( E_{\mu\nu} + \CA_{\mu\nu} \bigr) + \lambda \, \bar{\p} X^\mu \, \tau_\mu + \bar{\lambda} \, \p X^\mu \, \bar{\tau}_\mu \bigr) \Bigr]\,,
\end{align}
where we are in the real time with $\p = \p_t + \p_\sigma$ and $\bar{\p} = - \p_t + \p_\sigma$\,. Whether or not any additional spacetime torsional constraint is imposed does not change the following discussions.
The canonical momentum conjugate to $X^\mu$ is
\be \label{eq:cmom}
	P_\mu = E_{\mu\nu} \, \p_t {X}^\nu - \CA_{\mu\nu} \, \p_\sigma {X}{}^\nu + \frac{1}{2} \, \bigl( \lambda \, \tau_\mu - \bar{\lambda} \, \bar{\tau}_\mu \bigr)\,.
\ee
It follows that
\begin{subequations} \label{eq:lL}
\begin{align} 
	\lambda & = P + \CA_{\mu} \, \p_\sigma {X}^\mu, 
		&%
	E_\mu{}^{A'} \p_t X^\mu & = P_{A'} + E^\mu{}_{\!A'} \, \CA_{\mu\nu} \, \p_\sigma {X}{}^\nu, \\[2pt]
	\bar{\lambda} & = - \bar{P} - \bar{\CA}_{\mu} \, \p_\sigma {X}^\mu, 
\end{align}
\end{subequations}
where $P \equiv P_0 + P_1$\,, $\bar{P} \equiv P_0 - P_1$\,, $\CA_\mu \equiv \CA_{0\mu} + \CA_{1\mu}$\,, and $\bar{\CA}_{\mu} \equiv \CA_{0\mu} - \CA_{1\mu}$\,.
The Hamiltonian is given by
\be
	H = \frac{1}{2} \int_\Sigma d^2 \sigma \, \mathbb{X}^\text{I} \, \mathbb{G}_\text{IJ} \, \mathbb{X}^\text{J},
\ee
where 
\be
	\mathbb{X}^\text{I} = 
	\begin{pmatrix}
		P_\nu \\
		\p_\sigma X^\nu
	\end{pmatrix},
		\qquad%
	\mathbb{G}_\text{IJ} = 
	\begin{pmatrix}
		E^{\mu\nu} & L^\mu{}_\nu \\[2pt]
		M_\mu{}^\nu & N_{\mu\nu}
	\end{pmatrix},
\ee
and $\mathbb{G}_\text{IJ}$ is precisely the generalized metric in nonrelativistic string theory, with
\begin{subequations}
\begin{align}
	L^\mu{}_\nu & = E^{\mu\rho} \, \CA_{\rho\nu} - \epsilon^A{}_B \, \tau^\mu{}_A \, \tau_\nu{}^B,
		\qquad%
	M_\mu{}^\nu = - \CA_{\mu\rho} \, E^{\rho\nu} + \epsilon_A{}^B \, \tau_\mu{}^A \, \tau^\nu{}_{\!B}\,, \\[2pt]
	N_{\mu\nu} & = E_{\mu\nu} - \CA_{\mu\rho} \, E^{\rho\sigma} \, \CA_{\sigma\nu} - 2 \, \epsilon^A{}_B \, \tau^\rho{}_{A} \, \CA_{\rho(\mu} \, \tau_{\nu)}{}^B.
\end{align}
\end{subequations}
Taking the local field redefinition $P_\mu \equiv - \p_\sigma Y_\mu$\,, we find the dual coordinate $Y_\mu$ conjugate to string windings. The Hamiltonian is therefore manifestly invariant under T-dualities. It is also a straightforward exercise to show that the same $\mathbb{G}_\text{IJ}$ arises as a $U \rightarrow 0$ limit of the generalized metric in relativistic string theory. Moreover, one can also show that all the components in $\mathbb{G}_\text{IJ}$ are invariant under the Stueckelberg symmetries. Similar discussions in the context of double field theory can be found in \cite{Blair:2020gng, Ko:2015rha, Morand:2017fnv}. In the most general case, the spacetime geometry encoded by the generalized metric $\mathbb{G}_\text{IJ}$ depends on both the coordinates $X^\mu$ and $Y_\mu$\,, and nonlocal features due to string windings will become visible there. 

It is also interesting to apply the first two equations in \eqref{eq:lL} as a redefinition of $\lambda$ and $\bar{\lambda}$ in \eqref{eq:S}. This leads to
\begin{align} \label{eq:foaction}
\begin{split}
	S = - \frac{1}{4\pi\alpha'} \int_\Sigma d^2 \sigma \Bigl[ - \p_t X^\mu \, \p_t X^\nu \, E_{\mu\nu} & + \p_\sigma X^\mu \, \p_\sigma X^\nu \, \tilde{E}_{\mu\nu} + 2 \, \p_t X^\mu \, \p_\sigma X^\nu \tilde{\CA}_{\mu\nu} \\[2pt]
	& + P \, \bar{\p} X^\mu \, \tau_\mu - \bar{P} \, \p X^\mu \, \bar{\tau}_\mu \bigr) \Bigr]\,,
\end{split}
\end{align}
where
\be
	\tilde{E}_{\mu\nu} = E_{\mu\nu} - 2 \, \epsilon^A{}_B \, \tau^\rho{}_{A} \, \CA_{\rho(\mu} \, \tau_{\nu)}{}^B\,,
		\qquad
	\tilde{\CA}_{\mu\nu} = \CA_{\mu\nu} - \tau^\rho{}_A \, \CA_{\rho\nu} \tau_\mu{}^A\,.
\ee
This is a partially first-order formalism. We further perform a field redefinition with $P_A = - \p_\sigma Y_\mu \, \tau^\mu{}_A$\,, where $Y_\mu$ satisfies the orthogonality condition $\p_\sigma Y_\mu \, E^{\mu}{}_{A'} = 0$\,. This field redefinition contributes the path integral measure non-dynamically. Therefore, we are free to plug these field redefinitions directly into the action \eqref{eq:foaction}, which yields
\begin{align}
\begin{split}
	S = - \frac{1}{4\pi\alpha'} \int_\Sigma d^2 \sigma \Bigl[ - \p_t X^\mu \, \p_t X^\nu \, E_{\mu\nu} & + \p_\sigma X^\mu \, \p_\sigma X^\nu \, \tilde{E}_{\mu\nu} + 2 \, \p_t X^\mu \, \p_\sigma X^\nu \tilde{\CA}_{\mu\nu} \\[2pt]
	& - 2 \, \p_\sigma Y_\mu \, \bar{\p} X^\nu \, \tau^\mu \, \tau_\nu + 2 \, \p_\sigma Y_\mu \, \p X^\nu \, \bar{\tau}^\mu \, \bar{\tau}_\nu \bigr) \Bigr]\,.
\end{split}
\end{align}
This is in analogy with the Tseytlin's formalism in \cite{Tseytlin:1990nb} and metastring theory \cite{Freidel:2015pka}, but now in nonrelativistic string theory with only longitudinal monodromy.

\section{Nonrenormalization from Torsional Constraints} \label{app:nonre}

In this appendix, we show that, at $U = 0$\,, $\beta^U$ discussed in \S\ref{sec:bftd} vanishes at higher loop orders under the condition $T_{\mu\nu} = - \Omega_{[\mu} \tau_{\nu]}$ or $\overline{T}_{\mu\nu} = \Omega_{[\mu} \bar{\tau}_{\nu]}$ in \eqref{eq:dt0}.

For simplicity, we set $\alpha' = 1/(2\pi)$ in the following calculation. We focus on the $\lambda$- and $\bar{\lambda}$-dependent terms in the action \eqref{eq:SEA}, with $U = 0$\,,
\begin{align} \label{eq:Sll}
	S_\lambda = \frac{1}{2} \int_\Sigma d^2 \sigma \, \sqrt{h} \, \lambda \, \bar{\CD} X^\mu \, \tau_\mu [X]\,,
		\qquad
	S_{\bar{\lambda}} = \frac{1}{2} \int_\Sigma d^2 \sigma \, \sqrt{h} \, \bar{\lambda} \, \CD X^\mu \, {\bar{\tau}}_\mu [X]\,.
\end{align}
Since we already tuned $U = 0$\,, so the $\lambda \bar{\lambda}$ term is not included. Any quantum corrections to the $\lambda\bar{\lambda}$ operator necessarily involve vertices that arise from \eqref{eq:Sll}. Since the following calculation does not involve the dilaton term, it is sufficient to work with flat worldsheet, on which \eqref{eq:Sll} becomes
\begin{align} \label{eq:Sll2}
	S_\lambda = \frac{1}{2} \int_\Sigma d^2 \sigma \, \lambda \, \bar{\p} X^\mu \, \tau_\mu[X]\,,
		\qquad
	S_{\bar{\lambda}} = \frac{1}{2} \int_\Sigma d^2 \sigma \, \bar{\lambda} \, \p X^\mu \, {\bar{\tau}}_\mu[X]\,.
\end{align}
We now apply the background field method to compute quantum corrections from the interactions in \eqref{eq:Sll2} to the $\lambda\bar{\lambda}$ term in the effective action, with all loops taken into account.
For our purpose, it is sufficient to take a linear splitting of worldsheet fields, 
\be \label{eq:lex}
	X^\mu = X_0^\mu + \ell^\mu,
		\qquad
	\lambda = \lambda_0 + \rho\,,
		\qquad
	\bar{\lambda} = \bar{\lambda}_0 + \bar{\rho}\,,
\ee
where $X_0^\mu$\,, $\lambda_0$\,, and $\bar{\lambda}_0$ are classical fields that depend on the worldsheet coordinates, and $\ell^\mu$, $\rho$\,, and $\bar{\rho}$ are quantum fluctuations to be integrated out in the path integral. 

Consider a Feynman diagram $\Gamma$ (of any loop order) that contributes quantum corrections to the marginal $\lambda_0 \bar{\lambda}_0$ term in the effective action. This $\Gamma$ necessarily involves one vertex proportional to $\lambda_0$ and one vertex proportional to ${\bar{\lambda}}_0$\,. Focusing on the action terms pertain to the desired vertices that involve either $\lambda_0$ or ${\bar{\lambda}}_0$\,, and taking into account the expansions in \eqref{eq:lex}, we find
\begin{align} \label{eq:Slex}
	S_\lambda & = \frac{1}{2} \int_\Sigma d^2 \sigma \, \bigl( \lambda_0 + \rho \bigr)\, \bar{\p} \bigl( X_0^\mu + \ell^\mu \bigr) \, \sum_{k=0}^\infty \frac{1}{k!} \, \ell^{\nu^{}_1} \cdots \ell^{\nu^{}_k} \, \p_{\nu^{}_1} \cdots \p_{\nu^{}_k} \tau_\mu [X_0] \notag \\[2pt]
	& = \frac{1}{2} \int_\Sigma d^2 \sigma \, \lambda_0 \lc \sum_{k=1}^\infty \frac{\p_{\nu^{}_1} \cdots \p_{\nu^{}_k} \tau_\mu [X_0]}{(k+1)!} \, \bigl( k \, \bar{\p} \ell^\mu - \ell^\mu \bar{\p} \, \bigr) \, \bigl( \ell^{\nu^{}_1} \cdots \ell^{\nu^{}_k} \bigr) + \bar{\p} \ell^\mu \, \tau_\mu [X_0] \rc + \cdots \notag \\[2pt]
	& = \int_\Sigma d^2 \sigma \lc \lambda_0 \, \bar{\p} \ell^\mu \sum_{k=1}^\infty \frac{k \, \ell^{\nu_1} \cdots \ell^{\nu_k} \, \p_{\nu_1} \cdots \p_{\nu_{k-1}} T_{\nu^{}_k \, \mu}}{(k+1)!} + \tfrac{1}{2} \, \lambda_0 \, \bar{\p} \ell^\mu \, \tau_\mu[X_0] \rc + \cdots.
\end{align}
Here, $T_{\mu\nu} = \p_{[\mu} \tau_{\nu]} [X_0]$ and ``$\cdots$" denotes terms that are not relevant to the following discussion. These omitted terms include the ones that involve $\bar{\p} X_0^\mu$ or $\bar{\p} \lambda_0$ and the ones that do not contain $\lambda_0$\,.
The term $\lambda_0 \, \bar{\p} \ell^\mu \, \tau_\mu [X_0]$ in $S_\lambda$ is linear in the quantum fluctuation $\ell^\mu$. This term determines the equations of motion that the background fields satisfy. From \eqref{eq:Slex}, we observe that the vertex in $\Gamma$ that gives rise to the external $\lambda_0$ leg must be proportional to (derivatives of) $T_{\mu\nu}$\,. Applying the same reasoning to $S_{\bar{\lambda}}$ in \eqref{eq:Sll2}, we also conclude that $\Gamma$ is proportional to (derivatives of) ${\overline{T}}_{\mu\nu}$\,. Therefore,
\be
	\Gamma \propto \lambda_0 \, \bar{\lambda}_0 \, \p_{\gamma_1} \cdots \p_{\gamma_n} T_{\mu\rho} \, \p_{\kappa_1} \cdots \p_{\kappa_{\bar{n}}} \overline{T}_{\nu\sigma}\,.
\ee
Summing over all such Feynman diagrams evaluates the full quantum correction to the $\lambda \bar{\lambda}$ operator. This quantum correction includes contributions at all loop orders. The divergent part of the quantum correction is by power counting logarithmic divergent, and contributes the beta functional of $U[X]$\,. Recall that the physical value of $U$ has been fine tuned to zero. The final result has to be covariant, which implies that
\be \label{eq:betaUfull}
	\beta^U = \lambda_0 \, \bar{\lambda}_0 \sum_{n, \, \bar{n} = 0}^\infty \alpha^{\mu\nu\rho\sigma \, \gamma_1 \cdots \gamma_n \, \kappa_1 \cdots \kappa^{}_{\bar{n}}} \, \nabla_{\!\gamma^{}_1} \cdots \nabla_{\!\gamma_n} D_{[\mu} \tau_{\rho]} \, \nabla_{\!\kappa^{}_1} \cdots \nabla_{\!\kappa^{}_{\bar{n}}} D_{[\nu} \bar{\tau}_{\sigma]}\,.
\ee 
Here, $D_{[\mu} \tau_{\nu]}{}^A \equiv T_{\mu\nu}{}^A + \epsilon^A{}_B \, \Omega_{[\mu} \tau_{\nu]}{}^B$ and $\nabla_{\!\mu}$ is the covariant derivative associated with the string Newton-Cartan geometry. The detailed form of the coefficient $\alpha^{\mu\nu\rho\sigma\cdots}$ does not matter for our nonrenormalization argument. From \eqref{eq:betaUfull}, we conclude that $\beta^U$ is exactly zero when $D_{[\mu} \tau_{\nu]} = 0$ or $D_{[\mu} \bar{\tau}_{\nu]} = 0$ is satisfied. The same argument also shows that the finite part of the quantum corrections to the $\lambda \bar{\lambda}$ operator is zero.

In \S\ref{sec:netc}, we require the $Z$ ($\overline{Z}$) symmetry in the sigma model such that $D_{[\mu} \bar{\tau}_{\nu]} = 0$ ($D_{[\mu} \tau_{\nu]} = 0$) is imposed \emph{a priori}. This leads to the geometric constraints in \eqref{eq:gcsusy}. According to the discussions in this appendix, the $\lambda\bar{\lambda}$ operator is not generated (finitely nor divergently) at all loops when the geometric constraints \eqref{eq:gcsusy} are imposed. Therefore, imposing the $Z$ symmetry in the worldsheet QFT is sufficient for protecting the sigma model that describes nonrelativistic string theory from being deformed by the $\lambda \bar{\lambda}$ operator towards the full string theory.

\section{Symmetry Algebras} \label{app:sa}
    
In this appendix, we collect different symmetry algebras that have been referred to in the bulk of these notes. 
    
\subsection*{D.1. Poincar\'{e} Algebra} \label{app:pa}

In the Poincar\'{e} symmetry algebra that underlies the free action \eqref{eq:defaction}, the translational generator $\hat{P}_{M}$ and the Lorentz generator $\hat{M}_{MN}$ are now parametrized by $U_0$\,, with
\begin{subequations} \label{eq:decompogen}
\begin{align}
	\hat{P}_A & = U_0^{1/2} \, H_A\,, 
		&\!\!\!\!\!\!%
	\hat{M}_{AB} & = M \, \epsilon_{AB}\,, 
		\qquad\,\,\,\,\, 
	\hat{M}_{A'B'} = J_{A'B'}\,, \\[2pt]
	\hat{P}_{A'} & = P_{A'}\,,
		&\!\!\!\!\!\!%
	\hat{M}_{AA'} & = - \hat{M}_{A'A} = U_0^{-1/2} \, G_{AA'}\,.
\end{align} 
\end{subequations}
We will later consider the contraction $U_0 \rightarrow 0$\,. 
The Poincar\'{e} algebra is defined by the nonvanishing commutators,
\begin{subequations} \label{eq:paapp}
\begin{align}
	[\hat{P}^{}_L\,, \hat{M}^{}_{MN}] & = \eta^{}_{LM} \, \hat{P}^{}_{N} - \eta^{}_{LN} \, \hat{P}^{}_M\,, \\[2pt]
	[\hat{M}^{}_{KL}\,, \hat{M}^{}_{MN}] & = - \eta^{}_{KM} \, \hat{M}^{}_{LN} + \eta^{}_{LM} \, \hat{M}^{}_{KN} - \eta^{}_{LN} \, \hat{M}^{}_{KM} + \eta^{}_{KN} \, \hat{M}^{}_{LM}\,.
\end{align}
\end{subequations}
Plugging \eqref{eq:decompogen} into \eqref{eq:paapp} gives
\begin{subequations} \label{eq:unchanged}
\begin{align}
    [H_A\,, M] & = \epsilon_A{}^B \, H_B\,, &
    [H_A\,, G_{BA'}]  & = \eta^{}_{AB} \, P_{A'}\,, \\[2pt]
    [P_{A'}\,, J_{B'C'}] & = \delta_{A'B'} \, P_{C'} - \delta_{A'C'} \, P_{B'}\,, &
    [G_{AA'}\,, M] & = \epsilon_A{}^B \, G_{BA'}\,, \\[2pt]
    [G_{AA'}\,, J_{B'C'}] & = \delta_{A'B'} \, G_{AC'} - \delta_{A'C'} \, G_{AB'}\,, 
\end{align}
\vspace{-0.7cm}
\begin{align}
    \hspace{-1.75cm}[J_{A'B'}\,, J_{C'D'}] = \delta_{B'C'} \, J_{A'D'} - \delta_{A'C'} \, J_{B'D'} + \delta_{A'D'} \, J_{B'C'} - \delta_{B'D'} \, J_{A'C'} \,,
\end{align}
\end{subequations}
together with
\be \label{eq:GP}
	[G_{AA'}\,, P_{B'}] = U_0 \, \delta_{A'B'} \, H_A\,,
		\qquad
	[G_{AA'}\,, G_{BB'}] = - U_0 \, \bigl( \eta_{AB} \, J_{A'B'} + \delta_{A'B'} \, \epsilon_{AB}  \, M \bigr)\,,
\ee
where $L = (A, A')$\,. This is of course identical to the Poincar\'{e} algebra \eqref{eq:paapp}. For later discussions, it is useful to note the operator representation of different generators,
\begin{subequations}
\begin{align}
	\text{longitudinal translations} \qquad H_A & = \p_A \\[2pt]
	\text{transverse translations} \qquad P_{A'} & = \p_{A'} \\[2pt]
	\text{longitudinal Lorentz boost} \qquad M & = \epsilon^{AB} \, X_B \, \p_A \\[2pt]
	\text{Lorentz transformations} \qquad G_{AA'} & = X_A \, \p_{A'} - U_0 \, X_{A'} \, \p_A \label{eq:GAA'Lo} \\[2pt]
	\text{transverse rotations} \qquad J_{A'B'} & = X_{A'} \, \p_{B'} - X_{B'} \, \p_{A'}
\end{align}
\end{subequations}

Moreover, the original action \eqref{eq:defaction} is also invariant under the longitudinal dilatational symmetry, generated by the longitudinal dilatational generator $D$\,. 
The additional commutators that involve $D$ are
\be \label{eq:da}
	[D\,, H_A] = H_A\,,
		\qquad
	[G_{AA'}\,, D] = G_{AA'}\,.
\ee

Note that the commutators in \eqref{eq:unchanged} will not change in any of the following algebras that we are about to discuss.

\subsection*{D.2. String Galilei Algebra} \label{app:sga}

In the contraction $U_0 \rightarrow 0$\,, we find the string Galilei algebra, in which the commutators in \eqref{eq:unchanged} and \eqref{eq:da} remain unchanged, while both the commutators in 
\eqref{eq:GP} now vanish. %
Unlike \eqref{eq:GAA'Lo}, now, $G_{AA'} = X_A \, \p_{A'}$ is the string Galilei boost generator. This string Galilei algebra can be deformed back to the Poincar\'{e} algebra by turning on $U_0 \neq 0$\,.

\subsection*{D.3. String Newton-Cartan Algebra} \label{app:sNCa}

In \cite{Andringa:2012uz, Bergshoeff:2018yvt}, a larger symmetry subalgebra of the infinite-dimensional algebra \eqref{eq:globaltrnsf} than the string Galilei algebra has been realized in the string sigma model \eqref{eq:zeroUaction}. This subalgebra is referred to as the string Bargmann algebra \cite{Bergshoeff:2019pij, Brugues:2006yd, Brugues:2004an, Bergshoeff:2018vfn}. In addition to the unchanged commutators in \eqref{eq:unchanged}, we also have the noncentral extension generators $Z_A$ and $Y$ that satisfy the following nonvanishing commutators:
\begin{subequations} \label{eq:sba}
\begin{align}
    [G_{AA'}\,, P_{B'}] & = \delta_{A'\!B'} \, Z_A\,,
    	&%
    [Z_A\,, M] & = \epsilon_A{}^B \, Z_B\,, \label{eq:ZA1} \\
    [G_{AA'}\,, G_{BB'}] & = \delta_{A'\!B'} \, \epsilon_{AB} \, Y\,,
        &%
    [H_A\,, Y]  & = \epsilon_A{}^B \, Z_A \,, \label{eq:ZA2}
\end{align}
\end{subequations}
Note that $Y$ has to be added for the Lie algebra to be closed. Moreover, $Y$ can be further extended to be $Y_{AB}$ with the traceless condition $Y_A{}^A = 0$\,, which leads to the string Newton-Cartan algebra \cite{Harmark:2018cdl, Bergshoeff:2019pij}. This is the largest symmetry group that has been realized in the interacting theory defined by \eqref{eq:gc}. This further extension to include $Y_{AB}$ does not play any important role in our discussion. 

Imposing the symmetry transformations generated by the string Newton-Cartan algebra on the sigma model \eqref{eq:zeroUaction} as in \eqref{eq:trnsfG} leads to the geometric constraints in \eqref{eq:gc}. These constraints prohibit the torsional deformation $\lambda \bar{\lambda}$ from being generated at all loops. Since the longitudinal dilatational symmetry is not compatible with the required geometric constraints in \eqref{eq:gc}, the symmetry algebra is \emph{not} supplemented with any commutators that involve the dilatational generator. Unlike the string Galilei algebra, the extended algebras with a $Z_A$ generator is no longer deformable to the Poincar\'{e} algebra. 

\subsection*{D.4. Modified String Newton-Cartan Algebra} \label{app:msNCa}

It is possible to break half of the $Z_A$ symmetry while preserving the longitudinal Lorentz boosts. Define $Z \equiv Z_0 + Z_1$ and $\overline{Z} \equiv Z_0 - Z_1$\,, and setting $\overline{Z} = 0$ in the string Bargmann algebra, we find that \eqref{eq:ZA1} should be replaced with the nontrivial commutators,
\begin{subequations} \label{eq:Zsba}
\begin{align} 
    [G_{A'}\,, P_{B'}] & = \delta_{A'\!B'} \, Z\,,
    	&%
    [Z\,, M] & = Z\,, \\
    [G_{AA'}\,, G_{BB'}] & = \delta_{A'\!B'} \, \epsilon_{AB} \, Y\,,
        &%
    [H\,, Y]  & = Z \,,
\end{align}
\end{subequations}
while the remaining commutators in \eqref{eq:unchanged} are unchanged.
Here, $G_{A'} \equiv G_{0A'} + G_{1A'}$ and $H \equiv H_0 + H_1$\,. We only kept the nontrivial commutators here. These commutators, together with the ones in \eqref{eq:unchanged}, form a subalgebra of the string Bargmann algebra. One may also add in the $Y_{AB}$ extensions as in the string Newton-Cartan algebra. Moreover, the sigma model is also invariant under the symmetries generated by the longitudinal dilatation, which means that our symmetry algebra is now supplemented with the commutators in \eqref{eq:da}, together with 
$[Z\,, D] = Z$ and $[Y, D] = 2 \, Y$.

\newpage

\bibliographystyle{JHEP}
\bibliography{tdnst}

\end{document}